\documentclass[paper,notoc]{JHEP3}
\input epsf
\def\tpone#1{
\begin{picture}(180,100)(0,#1)
  \SetWidth{0.5}
  \SetColor{Black}
  \EBox(90,20)(150,80)
  \Photon(90,80)(30,80){1.5}{10} 
  \Photon(30,80)(30,20){1.5}{10}
  \Photon(30,20)(90,20){1.5}{10}
  \Photon(10,80)(30,80){1.5}{3}
  \Photon(10,20)(30,20){1.5}{3}
  \Photon(150,80)(170,80){1.5}{3}
  \Photon(150,20)(170,20){1.5}{3}
  \Text(60,88)[]{$D_{11}$}
  \Text(18,50)[]{$D_{12}$}
  \Text(60,12)[]{$D_{13}$}
  \Text(120,12)[]{$D_{14}$}
  \Text(162,50)[]{$D_{15}$}
  \Text(120,88)[]{$D_{16}$}
  \Text(102,50)[]{$D_{17}$}
  \Text(5,88)[]{$p_1$}
  \Text(5,12)[]{$p_2$}
  \Text(165,88)[]{$p_1$}
  \Text(165,12)[]{$p_2$}
\end{picture} 
}

\def\tptwo#1{
\begin{picture}(180,100)(0,#1)
  \SetWidth{0.5}
  \SetColor{Black}
  \EBox(30,20)(150,80)
  \Photon(90,80)(90,20){1.5}{10} 
  \Photon(10,80)(30,80){1.5}{3}
  \Photon(10,20)(30,20){1.5}{3}
  \Photon(150,80)(170,80){1.5}{3}
  \Photon(150,20)(170,20){1.5}{3}
  \Text(60,88)[]{$D_{21}$}
  \Text(18,50)[]{$D_{22}$}
  \Text(60,12)[]{$D_{23}$}
  \Text(120,12)[]{$D_{24}$}
  \Text(162,50)[]{$D_{25}$}
  \Text(120,88)[]{$D_{26}$}
  \Text(102,50)[]{$D_{27}$}
  \Text(5,88)[]{$p_2$}
  \Text(5,12)[]{$p_1$}
  \Text(165,88)[]{$p_2$}
  \Text(165,12)[]{$p_1$}
\end{picture} 
}

\def\tpthree#1{
\begin{picture}(180,100)(0,#1)
  \SetWidth{0.5}
  \SetColor{Black}
  \Line(30,20)(30,80)
  \Line(30,80)(150,80)
  \Line(30,20)(90,20)
  \Line(90,20)(150,80)
  \Photon(90,20)(150,20){1.5}{10}
  \Photon(150,20)(90,80){-1.5}{11}
  \Photon(10,80)(30,80){1.5}{3}
  \Photon(10,20)(30,20){1.5}{3}
  \Photon(150,80)(170,80){1.5}{3}
  \Photon(150,20)(170,20){1.5}{3}
  \Text(60,88)[]{$D_{31}$}
  \Text(18,50)[]{$D_{36}$}
  \Text(60,12)[]{$D_{33}$}
  \Text(120,12)[]{$D_{37}$}
  \Text(145,41)[]{$D_{32}$}
  \Text(120,88)[]{$D_{35}$}
  \Text(145,60)[]{$D_{34}$}  
  \Text(5,88)[]{$p_1$}
  \Text(5,12)[]{$p_2$}
  \Text(165,88)[]{$p_2$}
  \Text(165,12)[]{$p_1$}
\end{picture} 
}

\def\diagone#1{
\begin{picture}(145,80)(0,#1)
  \Gluon(5,10)(30,10){2.5}{3}
  \Gluon(5,70)(30,70){2.5}{3}
  \Gluon(30,70)(30,10){2.5}{9}
  \Gluon(30,10)(75,10){2.5}{8}
  \Gluon(30,70)(75,70){2.5}{8}
  \ArrowLine(75,10)(75,70)
  \ArrowLine(75,70)(120,40)
  \ArrowLine(120,40)(75,10)
  \DashLine(120,40)(140,40){3}
\end{picture} 
}

\def\diagtwo#1{
\begin{picture}(160,80)(0,#1)
  \Gluon(5,20)(30,20){2.5}{3}
  \Gluon(5,60)(30,60){2.5}{3}
  \Gluon(30,60)(85,10){-2.5}{10}
  \Gluon(30,60)(85,70){2.5}{8}
  \ArrowLine(30,20)(85,70)
  \ArrowLine(85,10)(30,20)
  \ArrowLine(85,70)(130,40)
  \ArrowLine(130,40)(85,10)
  \DashLine(130,40)(150,40){3}
\end{picture} 
}

\def\sdiagone#1{
\begin{picture}(145,80)(0,#1)
  \Gluon(5,10)(30,10){2.5}{3}
  \Gluon(5,70)(30,70){2.5}{3}
  \Gluon(30,70)(30,10){2.5}{9}
  \Gluon(30,10)(75,10){2.5}{8}
  \Gluon(30,70)(75,70){2.5}{8}
  \DashArrowLine(75,10)(75,70){2}
  \DashArrowLine(75,70)(120,40){2}
  \DashArrowLine(120,40)(75,10){2}
  \DashLine(120,40)(140,40){3}
\end{picture} 
}

\def\sdiagtwo#1{
\begin{picture}(145,80)(0,#1)
  \Gluon(5,10)(75,10){2.5}{10}
  \Gluon(5,70)(30,70){2.5}{3}
  \Gluon(30,70)(75,10){2.5}{9}
  \Gluon(30,70)(75,70){2.5}{7}
  \DashArrowLine(75,10)(75,70){2}
  \DashArrowLine(75,70)(120,40){2}
  \DashArrowLine(120,40)(75,10){2}
  \DashLine(120,40)(140,40){3}
\end{picture} 
}

\def\fdiagone#1{
\begin{picture}(95,80)(0,#1)
  \Gluon(5,10)(25,10){2.5}{3}
  \Gluon(5,70)(25,70){2.5}{3}
  \ArrowLine(25,10)(25,70)
  \ArrowLine(25,70)(70,40)
  \ArrowLine(70,40)(25,10)
  \DashLine(70,40)(90,40){3}
\end{picture} 
}

\def\fdiagtwo#1{
\begin{picture}(95,80)(0,#1)
  \Gluon(5,10)(25,10){2.5}{3}
  \Gluon(5,70)(25,70){2.5}{3}
  \ArrowLine(25,70)(25,10)
  \ArrowLine(70,40)(25,70)
  \ArrowLine(25,10)(70,40)
  \DashLine(70,40)(90,40){3}
\end{picture} 
}

\def\sdiagthree#1{
\begin{picture}(95,80)(0,#1)
  \Gluon(5,10)(25,10){2.5}{3}
  \Gluon(5,70)(25,70){2.5}{3}
  \DashArrowLine(25,10)(25,70){2}
  \DashArrowLine(25,70)(70,40){2}
  \DashArrowLine(70,40)(25,10){2}
  \DashLine(70,40)(90,40){3}
\end{picture} 
}

\def\sdiagfour#1{
  \begin{picture}(95,80)(0,#1)
  \Gluon(5,10)(25,10){2.5}{3}
  \Gluon(5,70)(25,70){2.5}{3}
  \DashArrowLine(25,70)(25,10){2}
  \DashArrowLine(70,40)(25,70){2}
  \DashArrowLine(25,10)(70,40){2}
  \DashLine(70,40)(90,40){3}
  \end{picture} 
}

\def\sdiagfive#1{
  \begin{picture}(90,60)(0,#1)
  \Gluon(5,10)(30,30){2.5}{5}
  \Gluon(5,50)(30,30){2.5}{5}
  \DashArrowArc(50,30)(20,-90,270){2}
  \DashLine(70,30)(85,30){3}
  \end{picture} 
}

\def\tad#1{
\begin{picture}(40,40)(0,#1)
  \SetWidth{0.5}
  \SetColor{Black}
  \Oval(20,20)(9,18)(-90)
\end{picture} 
}

\def\lightbubble#1{
\begin{picture}(70,30)(0,#1)
  \SetWidth{0.5}
  \SetColor{Black}
  \Line(10,14)(25,14)\Line(10,16)(25,16)
  \PhotonArc(40,15)(13.5,0,360){2.5}{12}
  \Line(55,14)(70,14)\Line(55,16)(70,16)
\end{picture} 
}

\def\bubble#1{
\begin{picture}(70,30)(0,#1)
  \SetWidth{0.5}
  \SetColor{Black}
  \Line(10,14)(25,14)\Line(10,16)(25,16)
  \CArc(40,15)(15,0,360)
  \Line(55,14)(70,14)\Line(55,16)(70,16)
\end{picture} 
}

\def\trianb#1{
\begin{picture}(80,47)(0,#1)
  \SetWidth{0.5}
  \SetColor{Black}
  \Line(5,24)(20,24)\Line(5,26)(20,26)
  \Line(20,25)(60,45)
  \Line(20,25)(60,5)
  \Line(60,45)(60,5)
  \Photon(60,5)(75,5){1.5}{2}
  \Photon(60,45)(75,45){1.5}{2}
\end{picture} 
}

\def\dt#1{
\begin{picture}(20,50)(0,#1)
  \SetWidth{0.5}
  \SetColor{Black}
  \Oval(10,42)(6,12)(-90)
  \Oval(10,18)(6,12)(-90)
\end{picture} 
}

\def\btone#1{
\begin{picture}(70,40)(0,#1)
  \SetWidth{0.5}
  \SetColor{Black}
  \Line(10,19)(25,19)\Line(10,21)(25,21)
  \PhotonArc(40,20)(13.5,0,360){2.5}{12}
  \Oval(59,27)(7,4)(-30)
  \Line(55,19)(70,19)\Line(55,21)(70,21)
\end{picture} 
}

\def\bttwo#1{
\begin{picture}(70,40)(0,#1)
  \SetWidth{0.5}
  \SetColor{Black}
  \Line(10,19)(25,19)\Line(10,21)(25,21)
  \CArc(40,20)(15,0,360)
  \Oval(59,27)(7,4)(-30)
  \Line(55,19)(70,19)\Line(55,21)(70,21)
\end{picture} 
}

\def\db#1{
\begin{picture}(100,40) (0,#1)
    \SetWidth{0.5}
    \SetColor{Black}
    \PhotonArc(35,30)(13.5,0,360){2.5}{12}
    \Line(5,29)(20,29)\Line(5,31)(20,31)
    \CArc(65,30)(15,0,360)
    \Line(80,29)(95,29)\Line(80,31)(95,31)
  \end{picture}
}

\def\ssonetwotwo#1{
\begin{picture}(80,40)(0,#1)
  \SetWidth{0.5}
  \SetColor{Black}
  \Line(7,19)(22,19)\Line(7,21)(22,21)
  \CArc(40,20)(18,0,360)
  \Photon(22,20)(58,20){2.5}{5}
  \Line(58,19)(73,19)\Line(58,21)(73,21)
  \Vertex(40,38){2.00}
  \Vertex(40,2){2.00}
\end{picture} 
}

\def\sstwoonetwo#1{
\begin{picture}(80,40)(0,#1)
  \SetWidth{0.5}
  \SetColor{Black}
  \Line(7,19)(22,19)\Line(7,21)(22,21)
  \CArc(40,20)(18,0,360)
  \Photon(22,20)(58,20){2.5}{5}
  \Line(58,19)(73,19)\Line(58,21)(73,21)
  \Vertex(40,38){2.00}
  \Vertex(40,20){2.00}
\end{picture} 
}

\def\sstad#1{
\begin{picture}(60,55)(0,#1)
  \SetWidth{0.5}
  \SetColor{Black}
  \CArc(30,35)(18,0,360)
  \Photon(12,35)(48,35){2.5}{5}
  \Vertex(30,53){2.00}
  \Text(31,59)[]{$\nu_1$}
  \Vertex(30,35){2.00}
  \Text(31,41)[]{$\nu_2$}
  \Vertex(30,17){2.00}
  \Text(31,11)[]{$\nu_3$}
\end{picture} 
}

\def\tria#1{
\begin{picture}(80,50)(0,#1)
  \SetWidth{0.5}
  \SetColor{Black}
  \Line(5,24)(20,24)\Line(5,26)(20,26)
  \Line(20,25)(60,45)
  \Line(20,25)(60,5)
  \CArc(28,25)(37,-31,31)
  \PhotonArc(85,25)(32,145,215){1.5}{5}
  \Photon(60,45)(75,45){1.5}{2}
  \Photon(60,5)(75,5){1.5}{2}
\end{picture} 
}

\def\triatwo#1{
\begin{picture}(80,50)(0,#1)
  \SetWidth{0.5}
  \SetColor{Black}
  \Line(5,24)(20,24)\Line(5,26)(20,26)
  \Line(20,25)(60,45)
  \Line(20,25)(60,5)
  \CArc(28,25)(37,-31,31)
  \PhotonArc(85,25)(32,145,215){1.5}{5}
 \CCirc(43,25){3}{Black}{Black}
  \Photon(60,45)(75,45){1.5}{2}
 \Photon(60,5)(75,5){1.5}{2}
\end{picture} 
}

\def\triathree#1{
\begin{picture}(80,50)(0,#1)
  \SetWidth{0.5}
  \SetColor{Black}
  \Line(5,24)(20,24)\Line(5,26)(20,26)
  \Line(20,25)(60,45)
  \Line(20,25)(60,5)
  \CArc(28,25)(37,-31,31)
  \PhotonArc(85,25)(32,145,215){1.5}{5}
  \Photon(60,45)(75,45){1.5}{2}
  \Photon(60,5)(75,5){1.5}{2}
  \Vertex(65,28){2.0}
  \Vertex(65,22){2.0}
\end{picture} 
}

\def\dtria#1{
\begin{picture}(120,50)(0,#1)
  \SetWidth{0.5}
  \SetColor{Black}
  \Line(5,24)(20,24)\Line(5,26)(20,26)
  \Photon(20,25)(60,45){1.5}{5}
  \Photon(20,25)(60,5){1.5}{5}
  \Line(60,45)(100,25)
  \Line(60,5)(100,25)
  \Line(60,45)(60,5)
  \Line(100,24)(115,24)\Line(100,26)(115,26) 
\end{picture} 
}

\def\dtriabound#1{
  \begin{picture}(70,50)(0,#1)
  \SetWidth{0.5}
  \SetColor{Black}
  \Line(5,24)(20,24)\Line(5,26)(20,26)
  \PhotonArc(36,25)(16.5,0,360){2.5}{12}
  \Photon(36,42)(36,8){1.5}{4}
  \Line(53,24)(68,24)\Line(53,26)(68,26)
  \end{picture}
}

\def\tritad#1{
\begin{picture}(80,50)(0,#1)
  \SetWidth{0.5}
  \SetColor{Black}
  \Line(5,24)(20,24)\Line(5,26)(20,26)
  \Oval(20,35)(10,5)(0)
  \Line(20,25)(60,45)
  \Line(20,25)(60,5)
  \Line(60,45)(60,5)
  \Photon(60,45)(75,45){1.5}{2}
  \Photon(60,5)(75,5){1.5}{2}
\end{picture} 
}

\def\mpfour#1{
\begin{picture}(70,50) (0,#1)
    \SetWidth{0.5}
    \SetColor{Black}
    \CArc(35,25)(20,0,360)
    \Line(0,24)(15,24)\Line(0,26)(15,26)
    \Photon(45,43)(60,43){1.5}{2}
    \Photon(45,7)(60,7){1.5}{2}
    \PhotonArc(17,-2)(28,19,92){1.5}{6}
  \end{picture}
}

\def\glasses#1{
\begin{picture}(100,40) (0,#1)
    \SetWidth{0.5}
    \SetColor{Black}
    \CArc(35,20)(15,0,360)
    \Line(5,19)(20,19)\Line(5,21)(20,21)
    \CArc(65,20)(15,0,360)
    \Line(80,19)(95,19)\Line(80,21)(95,21)
  \end{picture}
}

\def\tribub#1{
\begin{picture}(100,50)(0,#1)
  \SetWidth{0.5}
  \SetColor{Black}
  \Line(5,24)(20,24)\Line(5,26)(20,26)
  \CArc(30,25)(10,0,360)
  \Line(40,25)(80,45)
  \Line(40,25)(80,5)
  \Line(80,45)(80,5)
  \Photon(80,45)(95,45){1.5}{2}
  \Photon(80,5)(95,5){1.5}{2}
\end{picture} 
}

\def\mpsix#1{
\begin{picture}(95,50)(0,#1)
  \SetWidth{0.5}
  \SetColor{Black}
  \Line(5,24)(20,24)\Line(5,26)(20,26)
  \Line(20,25)(60,45)
  \Line(20,25)(60,5)
  \Photon(60,45)(60,5){1.5}{6}
  \Line(60,45)(75,45)
  \Line(60,5)(75,5)
  \Line(75,45)(75,5)
  \Photon(75,5)(90,5){1.5}{2}
  \Photon(75,45)(90,45){1.5}{2}
\end{picture} 
}

\def\mpfive#1{
\begin{picture}(95,50)(0,#1)
  \SetWidth{0.5}
  \SetColor{Black}
  \Line(5,24)(20,24)\Line(5,26)(20,26)
  \Line(20,25)(60,45)
  \Line(20,25)(60,5)
  \Photon(60,45)(60,5){1.5}{6}
  \Line(60,45)(75,45)
  \Line(60,5)(75,45)
  \Photon(75,45)(90,45){1.5}{2}
  \Photon(60,5)(75,5){1.5}{2}
\end{picture} 
}

\def\xtria#1{
\begin{picture}(100,50)(0,#1)
  \SetWidth{0.5}
  \SetColor{Black}
  \Line(5,24)(20,24)\Line(5,26)(20,26)
  \Line(20,25)(50,45)
  \Line(20,25)(50,5)
  \Line(50,45)(80,45)
  \Line(50,5)(80,45)
  \Photon(50,5)(80,5){1.5}{4}
  \Photon(50,45)(80,5){1.5}{6}
  \Photon(80,5)(95,5){1.5}{2}
  \Photon(80,45)(95,45){1.5}{2}
\end{picture} 
}

\usepackage{cite}
\usepackage{epsfig}
\usepackage{amssymb}
\usepackage{amsmath}
\usepackage[active]{srcltx}
\usepackage{axodraw}

\newcommand{\bea}{\begin{eqnarray}}
\newcommand{\eea}{\end{eqnarray}}
\newcommand{\bean}{\begin{eqnarray*}}
\newcommand{\eean}{\end{eqnarray*}}

\def\beq{\begin{equation}}
\def\be{\begin{equation}}
\def\eeq{\end{equation}}
\def\ee{\end{equation}}

\def\vep{\varepsilon}

\def\EPS{\epsilon}
\def\br(#1,#2){\left\langle#1#2\right\rangle}
\def\sq(#1,#2){\left[#1#2\right]}
\def\s(#1,#2){s_{#1 #2}}
\def\t(#1,#2,#3){s_{#1 #2 #3}}

\def\mtop{m_t}
\def\ud{\mathrm{d}}
\def\Cl(#1,#2){\mathrm{Cl}_{#1}(#2)}
\def\LS(#1,#2,#3){\mathrm{Ls}_{#1}^{(#2)}(#3)}
\def\LSC(#1,#2,#3){\mathrm{Lsc}_{#1,#2}(#3)}
\def\LSLSC(#1,#2,#3,#4){\mathrm{LsLsc}_{#1,#2,#3}(#4)}
\def\Li(#1,#2){\mathrm{Li}_{#1}(#2)}

\def\Label#1{\label{#1}%
  \smash{\hbox to0pt{\raise1ex\hbox{\tiny[#1]}\hss}}}

\preprint{{\tt CERN-PH-TH-2006-235}}

\title{Two-loop amplitudes and master integrals for the production of 
a Higgs boson via a massive quark and a scalar-quark loop}

\author{Charalampos Anastasiou$^1$  Stefan Beerli$^2$ Stefan Bucherer$^2$
 Alejandro Daleo$^2$ and Zoltan Kunszt$^2$\\
{\small 
$^{1}$TH Unit, PH Department CERN, 
CH-1211 Geneva 23\,, Switzerland\\
$^{2}$Institute for Theoretical Physics,
 ETH, 
CH-8093, Zurich\,, Switzerland\\ 
E-Mail: \email{babis@cern.ch}, \email{sbeerli@itp.phys.ethz.ch},
\email{stefabu@itp.phys.ethz.ch}, 
 \email{adaleo@itp.phys.ethz.ch},  
\email{kunszt@itp.phys.ethz.ch}}
}

\abstract{
We compute all two-loop master integrals  
which are required for the evaluation of next-to-leading 
order QCD corrections in Higgs boson production via gluon 
fusion. Many two-loop  amplitudes for $2 \to 1$ processes 
in the Standard Model and beyond  can be expressed in terms of these 
integrals using automated reduction techniques. 
These integrals also form a subset of the master integrals  for more 
complicated $2 \to 2$ amplitudes with massive propagators in the 
loops.  As a first application, we evaluate the two-loop amplitude 
for Higgs boson production in gluon fusion via a massive quark.  Our 
result is the first independent check of the  
calculation of Spira, Djouadi, Graudenz and Zerwas. 
We also present for the first time
the two-loop amplitude for $gg \to h$ via a massive squark. 
     }

\begin{document}

\allowdisplaybreaks[3]

\section{Introduction}
\label{sec:intro}

Next-to-leading (NLO) QCD corrections are important for a wide 
range of processes at the LHC.
Recently, there have been new breakthroughs 
in developing efficient techniques for one-loop amplitudes, 
e.g. ~\cite{pittau}. 
It is realistic that these theoretical advances will improve the 
data analysis for many interesting observables.   
However, processes which cannot occur with tree-level interactions at leading 
order,  require two-loop rather than one-loop amplitudes at NLO.

Loop induced processes should be  rather 
sensitive to new physics (see for example ~\cite{manohar}). It is possible  
that we revisit their NLO corrections  several times in the future, 
in order to include new types of particle interactions in the loops.  
In this paper, we study a basic LHC process of this kind:  
the production of a Higgs boson in gluon fusion. 

The two-loop QCD corrections for $gg \to H$ in the minimal 
Standard Model and its two-Higgs-doublet extensions have already been  
computed in~\cite{Spira:1995rr}.  In this calculation, 
the mass effects of the quark coupled to the Higgs boson were also 
fully accounted for. The impact of the NLO correction is 
striking; for example, the SM Higgs boson production 
cross-section increases by more than $70\%$.  

This calculation has never been verified in the 
literature, except in well known limits such as 
within the heavy top-quark approximation~\cite{Dawson:1990zj}. 
It is thus important to perform an independent computation. 
Our main 
objective, however, 
is to automatize the evaluation of the two-loop amplitude in QCD 
for this and other processes with similar distribution of massive particles 
in the loops. This is essential for future applications   
in gluon fusion and other processes for a variety of 
extensions of the Standard Model.

In 2001 Laporta introduced  a new algorithm for an automated 
reduction of multi-loop integrals to master integrals~\cite{laporta}. 
A parallel rapid development of the Mellin-Barnes method~\cite{smirnov,tausk,tausk1,ourmb,czakonmb}, 
the differential equation method~\cite{diffeqs}, 
and sector decomposition~\cite{sector1,sector2,muon}, 
yielded robust technology for computing master integrals. 
As a result, two-loop 
calculations for three and four point functions with more than 
one scale are now tractable.  

In this paper we study the two-loop integrals which are required 
in gluon fusion processes.  We apply the algorithm of Laporta, using 
the package AIR~\cite{air} and an independent MATHEMATICA 
implementation~\cite{ale-reduction}, to perform a reduction to master 
integrals. Some master integrals were already known in the 
literature~\cite{Bonciani:2003te,Bonciani:2003hc,Davydychev:2003mv,Fleischer:2004vb,Davydychev:2000na,
Broadhurst:1993mw,Kalmykov:2000qe}. We have recomputed them using the method of 
differential equations~\cite{diffeqs}. We present here for the first time the remaining master integrals, 
including the most complicated scalar cross-triangle. The same master integrals enter the evaluation of 
two-loop amplitudes of more complicated $2 \to 2$ processes, such as 
heavy quark pair production.   

Our results are first given as an expansion in the 
dimension parameter $\epsilon=(4-d)/2$ in terms of  
harmonic polylogarithms. We present the series coefficients through 
the order where harmonic polylogarithms with transcendentality four 
appear.
The polylogarithms are real valued in a non-physical kinematic 
region corresponding to imaginary center of mass energy $s <0$.
Then we perform explicitly the analytic continuation in the physical regions
below and above the threshold corresponding to two 
on-shell heavy particles in the loops: $s = 4 m_t^2$.   
For $s > 4 m_t^2$ the analytic 
continuation proceeds as in~\cite{analcont1, Remiddi:1999ew}.  For $s< 4m_t^2$, we find the 
analytic continuation of harmonic polylogarithms using the procedure described 
in~\cite{Davydychev:2003mv}. 
Interestingly, the result for the master 
integrals is in general simpler than the analytic continuation for individual 
harmonic polylogarithms.  

As a first application, we compute the two-loop amplitude for $gg\to h$ 
in the Standard Model.  We have compared our result with the 
expression in ~\cite{Harlander:2005rq},
in the non-physical region. The result of ~\cite{Harlander:2005rq} was 
derived by series expanding the integral 
representation of the  two-loop amplitude from ~\cite{Spira:1995rr} 
in a kinematic variable,  
and mapping the expansion to a carefully chosen ansatz.  Our result agrees 
with ~\cite{Harlander:2005rq}. This is the first independent check of the 
two-loop amplitude in~\cite{Spira:1995rr}.

We present here a new result for the two-loop amplitude $gg \to h$ via 
scalar-quarks. In the heavy squark limit, our result agrees 
with~\cite{Dawson:1996xz}. With a completed setup for the reduction procedure 
and the expressions for the master integrals, the evaluation 
of this new result is fully automated. Other amplitudes with different 
particle content in the loops can be obtained easily. We note that preliminary numerical 
results for the NLO K-factor for the squark case 
have been presented in~\cite{hp2spira}.

\section{Reduction of the amplitudes}
\label{sec:reduction}
We consider QCD
virtual corrections to the process 
\begin{equation}
g+g\rightarrow h
\end{equation}
with just one massive particle running in the loops. 
Some typical Feynman diagrams are shown in 
Figure \ref{fig:somediagrams} for the cases 
of a heavy quark and a squark in the loop. 

\FIGURE[h]{
\begin{tabular}{cc}
\diagone{37}&\diagtwo{37}\\
&\\
&\\
\multicolumn{2}{c}{(a)}\\
\sdiagone{37}&\sdiagtwo{37}\\
&\\
&\\
\multicolumn{2}{c}{(b)}
\end{tabular}
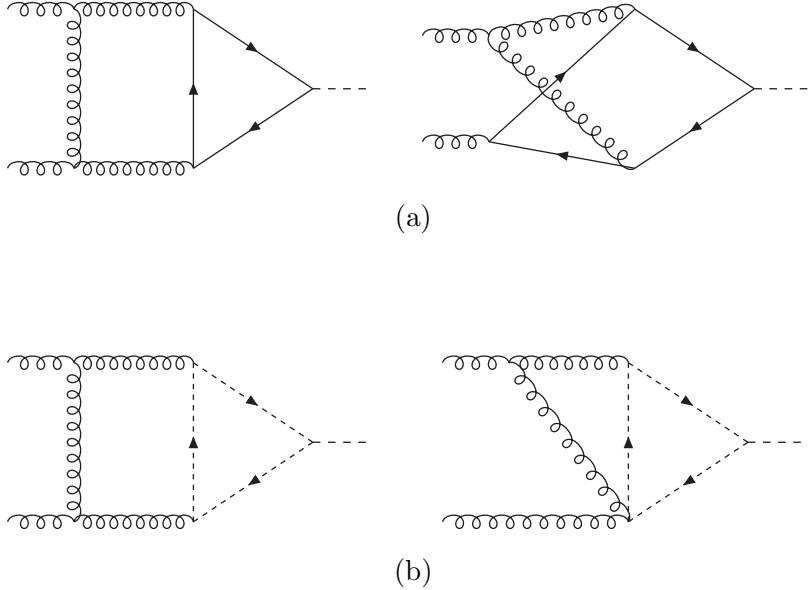
\caption{Typical Feynman diagrams in the two loop contributions to $gg\rightarrow H$ 
with (a) a heavy fermion in the loop, (b) a heavy scalar in the loop 
.}
\label{fig:somediagrams}
}

\FIGURE[h]{
\begin{tabular}{cc}
\tpone{47}&\tptwo{47}\\
&\\
&\\
$\mbox{TP}_1$&$\mbox{TP}_2$\\
\multicolumn{2}{c}{\tpthree{47}}\\
&\\
&\\
\multicolumn{2}{c}{$\mbox{TP}_3$}
\end{tabular}
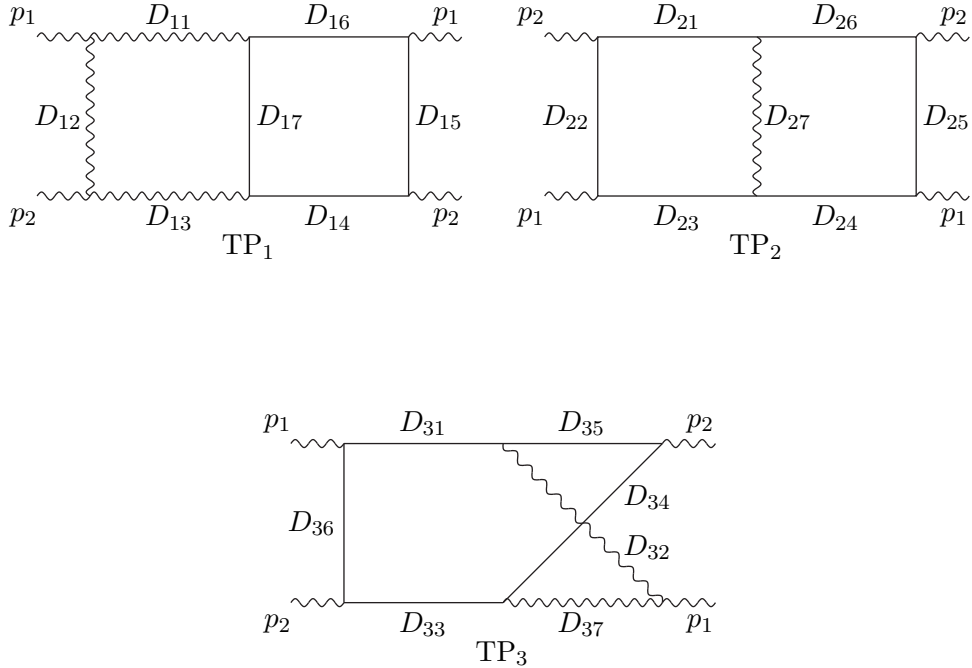
\caption{Master topologies. Wavy lines denote massless particles both external and internal. 
Internal massive lines are denoted by single straight lines whereas the double line denotes a 
massive external particle.
}\label{fig:topologies}
}
It is  convenient to project the two-loop amplitudes 
onto scalar form factors.
In this way we are left only with loop integrals involving scalar products 
in the numerator.
The scalar integrals can be classified into 
topologies according to their denominators. 
In $gg\rightarrow h$ we find diagrams with at most six propagators,
of which five can be massive. In the numerators, however, one can find 
seven independent scalar products. 
The irreducible scalar product can be dealt with by introducing  
an additional propagator, which  is raised  to  
negative powers in the expressions for the physical amplitudes. 
After the introduction  of the auxiliary propagators, 
all scalar integrals belong to subtopologies of 
the three topologies, shown in 
Figure \ref{fig:topologies}, 
with denominators: 
\begin{equation}
\begin{array}{rclccrclccrcl}
\multicolumn{3}{c}{\mbox{TP}_1}&&&\multicolumn{3}{c}{\mbox{TP}_2}&&&\multicolumn{3}{c}{\mbox{TP}_3}\\
D_{11} &=& k^2&&&                 D_{21} &=& k^2-m_t^2&&&            D_{31} &=& k^2 -m_t^2 \\
D_{12} &=& (k+p_1)^2&&&           D_{22} &=& (k+p_2)^2-m_t^2&&&       D_{32} &=& (k-l-p_1)^2\\
D_{13} &=& (k+p_{12})^2&&&        D_{23} &=& (k+p_{12})^2-m_t^2&&&    D_{33} &=& (k+p_{12})^2-m_t^2\\
D_{14} &=& (l + p_{12})^2-m_t^2&&&  D_{24} &=& (l + p_{12})^2-m_t^2&&&  D_{34} &=& (l + p_{12})^2-m_t^2 \\    
D_{15} &=& (l+p_1)^2-m_t^2&&&       D_{25} &=& (l+p_2)^2-m_t^2&&&       D_{35} &=& (l+p_1)^2-m_t^2\\ 
D_{16} &=& l^2-m_t^2&&&             D_{26} &=& l^2-m_t^2&&&             D_{36} &=& (k+p_1)^2-m_t^2\\
D_{17} &=& (k-l)^2-m_t^2&&&         D_{27} &=& (k-l)^2&&&            D_{37} &=& (k-l)^2\,,
\end{array}
\end{equation}
where $p_{12}=p_1+p_2$, $m_t$
is the mass of the particle running in the loops and $k$ and $l$ are
the loop momenta.

The reduction to master integrals is done using integration by part identities
\cite{Tkachov:1981wb,Chetyrkin:1981qh} combined with the Laporta 
algorithm~\cite{laporta} in~\cite{air,ale-reduction}. 
We found 17 master integrals, which are shown in 
Figure \ref{fig:masterlist}.
It is possible to choose  a different basis of master integrals; 
the basis we choose is particularly convenient for the method of differential equations.

\FIGURE[h]{
\begin{tabular}{cccc}
\dt{37}&\db{27}&\btone{16}&\bttwo{17}\\
&&&\\
\tritad{23}&\glasses{17}&\tribub{23}&\\
\tria{23}&\triatwo{23}&\triathree{23}&\dtria{23}\\
&&&\\
\ssonetwotwo{17}&\sstwoonetwo{17}&&\\
\mpfour{23}&\mpfive{23}&\mpsix{23}&\xtria{23}\\
&&&\\
&&&\\
\end{tabular}
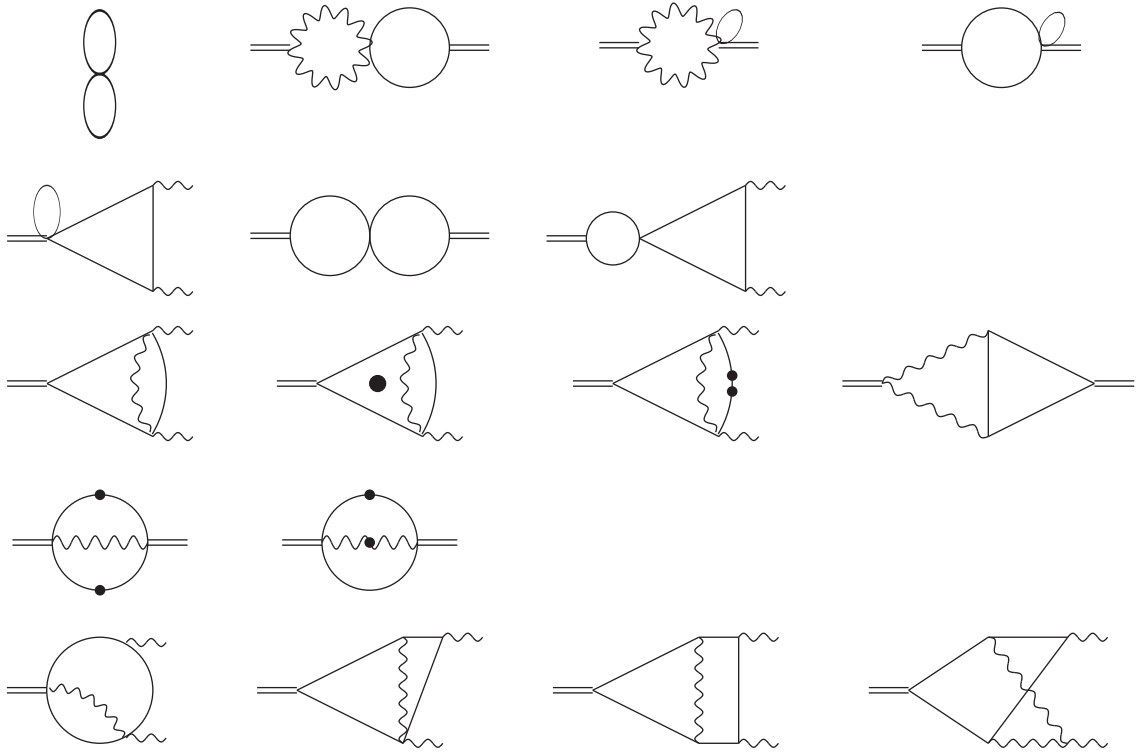
\caption{Set of master integrals. The conventions for the lines are as in Figure \ref{fig:topologies}. 
Each dot on a propagator line denotes an additional power of the propagator
in the denominator. The diagram with a big dot contains
a numerator, it is defined in the following section.}\label{fig:masterlist}
}

The master integrals in the first two lines of
Figure~\ref{fig:masterlist} are  products
of known one-loop integrals~\cite{Bonciani:2003te,Davydychev:2003mv}. 
The master integrals in the third, fourth and
fifth line in Figure \ref{fig:masterlist} are non-factorizable.
Integrals in the third and fourth line were calculated already in 
\cite{Bonciani:2003hc}\footnote{Our results 
fully agree with the results quoted in this reference taken from 
the electronic file in 
{\tt http://pheno.physik.uni-freiburg.de/bonciani/}. The printed version 
contains several typographical mistakes.} 
and \cite{Broadhurst:1993mw,Davydychev:2000na,Davydychev:2003mv}.
respectively. The double triangle, last diagram in the third line
was calculated in \cite{Broadhurst:1993mw,Kalmykov:2000qe,Davydychev:2000na}.
Also the six propagators triangle - third diagram in the last
line of Figure \ref{fig:masterlist} - has been calculated 
in~\cite{Fleischer:2004vb}.

\section{Master integrals}
\label{sec:master}

We computed all master integrals using the differential 
equation method \cite{diffeqs,Kotikov:1990kg,Kotikov:1991hm,
Kotikov:1991pm,Remiddi:1997ny,Caffo:1998yd}. 
The natural variable to express the results is
\beq\label{eq:xtau}
x = \frac{\sqrt{1-\tau}-1}{\sqrt{1-\tau}+1} + i\vep \qquad \textrm{where} \qquad \tau
= \frac{4 {\mtop}^2}{s} \,,
\eeq
with $s=(p_1+p_2)^2$. The variable $x$ is real valued in the space-like region 
($s<0$) and in the physical region above threshold 
($s > 4\,m_t^2$). Below threshold $x$ lies on the unit circle in the complex 
plane. 
In that region, we introduce the variable $\theta$ such that $x =
e^{i\theta}$.
For quick reference, in Table \ref{tab:kinematics} we list the domain of 
each variable in the different kinematical regions. 
\TABLE[h]{
\begin{tabular}{l|c|c|c}
{\bf Region} & $s$ & $\tau$ & $x$  \\
\hline
space-like & $-\infty < s < 0$ & $0 > \tau > -\infty$ & $0 < x < 1$ \\
\hline
below threshold & $0 < s < 4 {\mtop}^2$ & $\infty > \tau > 1$ & $x = e^{i\theta}$ with $0<\theta<\pi$ \\
\hline
above threshold & $4 {\mtop}^2 < s < \infty$ & $1 > \tau > 0$ & $-1 < x < 0$ \\
\end{tabular}
\caption{Domain spanned by the variables in the different kinematical regions}\label{tab:kinematics} 
}

The dependence on $x$ of the master integrals is determined by solving 
the  associated differential equations.
These are  obtained by taking the derivative with
respect to $x$ of the loop integrals and exchanging the order of the differential operator and
the loop integration. In this way, the derivative of a given master integral
is expressed in terms of scalar integrals which can again be reduced 
to masters. 
Applying this procedure to all integrals in our master basis, 
we derive  a closed system of  differential equations.

We solve the differential equations order by order in powers of $\epsilon$. 
Only integrations with kernels
$1/x$, $1/(1-x)$ and $1/(1+x)$ are required,  
and the solutions can be written exclusively in 
terms of  harmonic polylogarithms~\cite{Remiddi:1999ew}.

To fully determine the solution of the differential equations, 
we require the  value of the master integrals at a certain kinematic point. 
The value at $x=1$ is very easy to obtain. 
This limit corresponds to setting the external momenta to 
zero $(s=0)$. All non-factorizable master integrals  
$\mathrm{MI}^{(\mathrm{NF})}$ collapse to a vacuum sunset 
diagram with extra powers of propagators: 
\beq
\lim_{x \to 1} \mathrm{MI}^{(\mathrm{NF})} = \sstad{32}.
\eeq
{}\\
The exponent $\nu_2$ corresponds to the number of massless
propagators in the integral whereas $\nu_1 + \nu_3$ is the number of 
massive propagators. One can easily compute: 
\beq
\label{eq:sstad}
\sstad{32} = (-1)^{\nu_{123}}\, {\mtop}^{2\,(d-\nu_{123})} 
\frac{\Gamma\left(\nu_{123}-d\right) \Gamma\left(\nu_{12}-\frac{d}{2}\right)
\Gamma\left(\nu_{23}-\frac{d}{2}\right) 
\Gamma\left(\frac{d}{2} - \nu_2\right)}{
\Gamma(\nu_1)\Gamma(\nu_3)\Gamma\left(\frac{d}{2}\right) 
\Gamma\left(\nu_{13}+2\,\nu_2 - d \right)  } \ .
\eeq
{}\\
We have observed that one could fix the solution of the 
differential equations by requiring simply that the $x \to 1$ limit 
is finite, since the homogeneous solutions usually
diverge at $x=1$. The explicit formula for the 
limit $x =1$  in Eq.~\ref{eq:sstad} was then an additional consistency 
check of our calculation. 

In one master integral, the $x=1$ limit does not commute 
with the expansion around $\epsilon =0$, due  to a collinear singularity 
as  $s$ vanishes.  For  this master integral, we have used 
the massless limit $x \to 0$, which is well behaved:
\beq
\lim_{x\to 0} { \SetScale{0.9} \dtria{20}} = { \SetScale{0.9} \dtriabound{20}}
\ ,
\eeq
with
\beq
{ \SetScale{0.9} \dtriabound{20}} = (-s)^{-1-2\,\EPS}
\left(\frac{\Gamma(1+\EPS)}{(1-\EPS)}\right)^2 \left(-6\, \zeta(3) -
  \EPS\,  \frac{\pi^4}{10} + \mathcal{O}\left(\EPS^2\right) \right)
\eeq
{}\\
In the following sections we shall present our results in the space-like 
and $0<s < 4 m_t^2$ regions, and  describe the analytic continuation
procedure. We have performed several checks on our results 
for the master integrals. We have verified that our expressions satisfy 
the differential equations before and  after analytic continuation. 
We have also computed all master integrals numerically from their Feynman 
parameterization, by using sector decomposition~\cite{sector1,sector2,muon}. 
Our analytic expressions agree fully with the direct numerical evaluation.

\section{Results in the region $s < 0$}
\label{sec:1result}
We now present the results for the master 
integrals in the space-like region.  
We give the definition of the master integrals in terms of 
the propagators listed in section \ref{sec:reduction}, and 
their $\epsilon$-expansion in terms of harmonic polylogarithms.

\allowdisplaybreaks[2] 

\subsection{One loop integrals}
\begin{align}
{ \SetScale{0.9} \tad{17}} &= \int \!\! \frac{d^dk}{i
  \pi^{d/2}}\frac{1}{k^2-\mtop^2 +i\varepsilon}=\frac{\Gamma(1+\epsilon)}{1-\epsilon} \, (\mtop^2)^{- \epsilon
  +1} \cdot \frac{1}{\epsilon}
\end{align}
\begin{align}
{ \SetScale{0.9} \lightbubble{12}} &= \int  \!\!
\frac{d^dk}{i\pi^{d/2}}\frac{1}{D_{11} D_{13}}=\frac{\Gamma(1+\epsilon)}{1-\epsilon} \,
(\mtop^2)^{- \epsilon} \cdot \frac{1}{\epsilon} \frac{\Gamma(2-\epsilon) \Gamma(1-\epsilon)}{\Gamma(2-2\epsilon)} \left( \frac{(1-x)^2}{x} -i\varepsilon \right)^{-\epsilon}
\end{align}

\begin{align}
{ \SetScale{0.9} \bubble{12} } &= \int \!\! \frac{d^dk}{i \pi^{d/2}}
\frac{1}{D_{21}D_{23}}  
=\frac{\Gamma(1+\epsilon)}{1-\epsilon} \, (\mtop^2)^{-\epsilon}\sum_{i=-1}^{3} \epsilon^{i} F_{\mathrm{mbub}}^{i} (x) + \mathcal{O}(\epsilon^{4})\end{align}
\begin{align}
F_{\mathrm{mbub}}^{-1}(x)\quad = \quad &1\hfill\\
F_{\mathrm{mbub}}^{0}(x)\quad = \quad &\frac{1}{1-x}\Big\{-x+(x+1) H(0;x)+1\Big\} \hfill\\
F_{\mathrm{mbub}}^{1}(x)\quad = \quad &\frac{1}{1-x}\Big\{\frac{1}{6} \left(-\pi ^2 x-12 x-\pi ^2+12\right)+(x+1) H(0;x)-2 (x+1) H(-1,0;x)    \nonumber\hfill \\ & 
+(x+1) H(0,0;x)\Big\} \hfill\\
F_{\mathrm{mbub}}^{2}(x)\quad = \quad &\frac{1}{1-x}\Big\{-\frac{1}{6} \pi ^2 (x+1)-2 ((2+\zeta (3)) x+\zeta (3)-2)    \nonumber\hfill \\ & 
+\frac{1}{3} \pi ^2 (x+1) H(-1;x)-\frac{1}{6} \left(-12+\pi ^2\right) (x+1) H(0;x)    \nonumber\hfill \\ & 
-2 (x+1) H(0,-1,0;x)-2 (x+1) H(-1,0;x)+(x+1) H(0,0;x)    \nonumber\hfill \\ & 
+4 (x+1) H(-1,-1,0;x)-2 (x+1) H(-1,0,0;x)+(x+1) H(0,0,0;x)\Big\} \hfill\\
F_{\mathrm{mbub}}^{3}(x)\quad = \quad &\frac{1}{1-x}\Big\{-\frac{1}{40} \pi ^4 (x+1)-\frac{1}{3} \pi ^2 (x+1)-2 ((4+\zeta (3)) x+\zeta (3)-4)    \nonumber\hfill \\ & 
+\frac{1}{3} \pi ^2 (x+1) H(0,-1;x)+\frac{1}{3} (x+1) H(-1;x) \left(\pi ^2+12 \zeta (3)\right)    \nonumber\hfill \\ & 
-\frac{1}{6} (x+1) \left(\pi ^2+12 (-2+\zeta (3))\right) H(0;x)-2 (x+1) H(0,0,-1,0;x)    \nonumber\hfill \\ & 
-2 (x+1) H(0,-1,0;x)-\frac{2}{3} \pi ^2 (x+1) H(-1,-1;x)    \nonumber\hfill \\ & 
+\frac{1}{3} \left(-12+\pi ^2\right) (x+1) H(-1,0;x)-\frac{1}{6} \left(-12+\pi ^2\right) (x+1) H(0,0;x)    \nonumber\hfill \\ & 
+4 (x+1) H(0,-1,-1,0;x)-2 (x+1) H(0,-1,0,0;x)+4 (x+1) H(-1,0,-1,0;x)    \nonumber\hfill \\ & 
+4 (x+1) H(-1,-1,0;x)-2 (x+1) H(-1,0,0;x)+(x+1) H(0,0,0;x)    \nonumber\hfill \\ & 
+4 (x+1) H(-1,-1,0,0;x)-8 (x+1) H(-1,-1,-1,0;x)    \nonumber\hfill \\ & 
+(x+1) H(0,0,0,0;x)-2 (x+1) H(-1,0,0,0;x)\Big\} \hfill
\end{align}

\begin{align}
{ \SetScale{0.9} \trianb{23} } &= \int \!\! \frac{d^dk}{i \pi^{d/2}}
\frac{1}{D_{21}D_{22}D_{23}} 
=\frac{\Gamma(1+\epsilon)}{1-\epsilon} \, (\mtop^2)^{-\epsilon-1}\sum_{i=0}^{2} \epsilon^{i} F_{\mathrm{mtri}}^{i} (x) + \mathcal{O}(\epsilon^{3})\end{align}
\begin{align}
F_{\mathrm{mtri}}^{0}(x)\quad = \quad &-\frac{x H(0,0;x)}{(x-1)^2}\hfill\\
F_{\mathrm{mtri}}^{1}(x)\quad = \quad &\frac{x}{(1-x)^2}\Big\{\frac{1}{6} \pi ^2 H(0;x)+2 H(0,-1,0;x)+H(0,0;x)+3 \zeta (3)    \nonumber\hfill \\ & 
-H(0,0,0;x)\Big\} \hfill\\
F_{\mathrm{mtri}}^{2}(x)\quad = \quad &\frac{x}{(1-x)^2}\Big\{-\frac{1}{3} \pi ^2 H(0,-1;x)+H(0;x) \left(-\frac{\pi ^2}{6}+2 \zeta (3)\right)-3 \zeta (3)+\frac{\pi ^4}{72}    \nonumber\hfill \\ & 
+2 H(0,0,-1,0;x)-2 H(0,-1,0;x)+\frac{1}{6} \pi ^2 H(0,0;x)-4 H(0,-1,-1,0;x)    \nonumber\hfill \\ & 
+2 H(0,-1,0,0;x)+H(0,0,0;x)-H(0,0,0,0;x)\Big\} \hfill
\end{align}

\subsection{Factorizable integrals}
\begin{align}
{ \SetScale{0.9} \dt{23}} &= \int \!\! \frac{d^dk}{i \pi^{d/2}} \int \!\!
\frac{d^dl}{i \pi^{d/2}}\frac{1}{D_{15}D_{17}}
={ \SetScale{0.9} \tad{17}}\times { \SetScale{0.9} \tad{17}}
\end{align}
\begin{align}
{ \SetScale{0.9} \db{23}} &= \int \!\! \frac{d^dk}{i \pi^{d/2}} \int \!\!
\frac{d^dl}{i \pi^{d/2}}\frac{1}{D_{11}D_{13}D_{14}D_{16}}
={ \SetScale{0.9} \lightbubble{12}}\times { \SetScale{0.9} \bubble{12}}
\end{align}

\begin{align}
{ \SetScale{0.9} \btone{17}} &= \int \!\! \frac{d^dk}{i \pi^{d/2}} \int \!\!
\frac{d^dl}{i \pi^{d/2}}\frac{1}{D_{11}D_{13}D_{16}}
={ \SetScale{0.9} \lightbubble{12}}\times { \SetScale{0.9} \tad{17}}
\end{align}

\begin{align}
{ \SetScale{0.9} \bttwo{17}} &= \int \!\! \frac{d^dk}{i \pi^{d/2}} \int \!\!
\frac{d^dl}{i \pi^{d/2}}\frac{1}{D_{14}D_{16}D_{17}}
={ \SetScale{0.9} \bubble{12}}\times { \SetScale{0.9} \tad{17}}
\end{align}

\begin{align}
{ \SetScale{0.9} \tritad{23}} &= \int \!\! \frac{d^dk}{i \pi^{d/2}} \int \!\!
\frac{d^dl}{i \pi^{d/2}}\frac{1}{D_{14}D_{15}D_{16}D_{17}}
={ \SetScale{0.9} \trianb{23}}\times { \SetScale{0.9} \tad{17}}
\end{align}

\begin{align}
{ \SetScale{0.9} \glasses{17}} &= \int \!\! \frac{d^dk}{i \pi^{d/2}} \int \!\!
\frac{d^dl}{i \pi^{d/2}}\frac{1}{D_{21}D_{23}D_{24}D_{26}}
={ \SetScale{0.9} \bubble{12}}\times { \SetScale{0.9} \bubble{12}}
\end{align}

\begin{align}
{ \SetScale{0.9} \tribub{23}} &= \int \!\! \frac{d^dk}{i \pi^{d/2}} \int \!\!
\frac{d^dl}{i \pi^{d/2}}\frac{1}{D_{21}D_{23}D_{24}D_{25}D_{26}}
={ \SetScale{0.9} \trianb{23}}\times { \SetScale{0.9} \bubble{12}}
\end{align}

\vspace{0.5cm}

\subsection{Three propagator integrals}
\begin{align}
{ \SetScale{0.9} \ssonetwotwo{15}} &= \int \!\! \frac{d^dk}{i \pi^{d/2}} \int \!\! \frac{d^dl}{i \pi^{d/2}}\frac{1}{D_{11}D_{14}^{2}D_{17}^{2}}\nonumber\\ 
&=\left(\frac{\Gamma(1+\epsilon)}{1-\epsilon}\right)^2 \, (\mtop^2)^{-2 \epsilon -1}\sum_{i=0}^{2} \epsilon^{i} F_{1}^{i} (x) + \mathcal{O}(\epsilon^{3})\end{align}
\begin{align}
F_{1}^{0}(x)\quad = \quad &-\frac{2 x H(0,0;x)}{(x-1)^2}\hfill\\
F_{1}^{1}(x)\quad = \quad &\frac{x}{(1-x)^2}\Big\{\frac{1}{3} \pi ^2 H(0;x)+12 H(0,-1,0;x)+4 H(0,0;x)+6 \zeta (3)    \nonumber\hfill \\ & 
-4 H(0,1,0;x)-6 H(0,0,0;x)+4 H(1,0,0;x)\Big\} \hfill\\
F_{1}^{2}(x)\quad = \quad &\frac{x}{(1-x)^2}\Big\{-2 \pi ^2 H(0,-1;x)-12 \zeta (3)+\frac{13 \pi ^4}{180}    \nonumber\hfill \\ & 
+\left(-\frac{2 \pi ^2}{3}+16 \zeta (3)\right) H(0;x)+\frac{2}{3} \pi ^2 H(0,1;x)-12 H(1;x) \zeta (3)    \nonumber\hfill \\ & 
+36 H(0,0,-1,0;x)-24 H(0,-1,0;x)+\left(-2+\pi ^2\right) H(0,0;x)-\frac{2}{3} \pi ^2 H(1,0;x)    \nonumber\hfill \\ & 
+8 H(0,1,0;x)-12 H(0,0,1,0;x)-72 H(0,-1,-1,0;x)+48 H(0,-1,0,0;x)    \nonumber\hfill \\ & 
+24 H(0,-1,1,0;x)+12 H(0,0,0;x)-24 H(1,0,-1,0;x)-8 H(1,0,0;x)    \nonumber\hfill \\ & 
+8 H(1,0,1,0;x)+24 H(0,1,-1,0;x)-20 H(0,1,0,0;x)-8 H(0,1,1,0;x)    \nonumber\hfill \\ & 
-14 H(0,0,0,0;x)+12 H(1,0,0,0;x)-8 H(1,1,0,0;x)\Big\} \hfill
\end{align}

\begin{align}
{ \SetScale{0.9} \sstwoonetwo{15}} &= \int \!\! \frac{d^dk}{i \pi^{d/2}} \int \!\! \frac{d^dl}{i \pi^{d/2}}\frac{1}{D_{11}^{2}D_{14}^{2}D_{17}}\nonumber\\ 
&=\left(\frac{\Gamma(1+\epsilon)}{1-\epsilon}\right)^2 \, (\mtop^2)^{-2 \epsilon -1}\sum_{i=-1}^{2} \epsilon^{i} F_{2}^{i} (x) + \mathcal{O}(\epsilon^{3})\end{align}
\begin{align}
F_{2}^{-1}(x)\quad = \quad &\frac{x H(0;x)}{x^2-1}\hfill\\
F_{2}^{0}(x)\quad = \quad &\frac{x}{(x-1)^3 (x+1)}\Big\{-2 H(0;x) (x-1)^2-\frac{1}{6} \pi ^2 (x-1)^2    \nonumber\hfill \\ & 
+(x-1) (5 x-3) H(0,0;x)-6 (x-1)^2 H(-1,0;x)    \nonumber\hfill \\ & 
+2 (x-1)^2 H(1,0;x)\Big\} \hfill\\
F_{2}^{1}(x)\quad = \quad &\frac{x}{(x-1)^3 (x+1)}\Big\{\frac{1}{3} (x-1) \left(6 \zeta (3) (4-7 x)+\pi ^2 (x-1)\right)    \nonumber\hfill \\ & 
+\pi ^2 (x-1)^2 H(-1;x)-\frac{1}{6} (x-1) \left(-6 x+\pi ^2 (5 x-3)+6\right) H(0;x)    \nonumber\hfill \\ & 
-\frac{1}{3} \pi ^2 H(1;x) (x-1)^2-6 (5 x-3) H(0,-1,0;x) (x-1)    \nonumber\hfill \\ & 
+12 (x-1)^2 H(-1,0;x)-2 (x-1) (5 x-3) H(0,0;x)    \nonumber\hfill \\ & 
+2 (x-1) (5 x-3) H(0,1,0;x)-4 (x-1)^2 H(1,0;x)    \nonumber\hfill \\ & 
+36 (x-1)^2 H(-1,-1,0;x)-24 (x-1)^2 H(-1,0,0;x)    \nonumber\hfill \\ & 
+(x-1) (13 x-7) H(0,0,0;x)-12 (x-1)^2 H(-1,1,0;x)    \nonumber\hfill \\ & 
+2 (x-1) (3 x-5) H(1,0,0;x)-12 (x-1)^2 H(1,-1,0;x)    \nonumber\hfill \\ & 
+4 (x-1)^2 H(1,1,0;x)\Big\} \hfill\\
F_{2}^{2}(x)\quad = \quad &\frac{x}{(x-1)^3 (x+1)}\Big\{-\frac{1}{360} (x-1) \left(60 \pi ^2 (x-1)+\pi ^4 (61 x-35)-1440 (7 x-4) \zeta (3)\right)    \nonumber\hfill \\ & 
+\pi ^2 (x-1) (5 x-3) H(0,-1;x)-2 (x-1)^2 H(-1;x) \left(\pi ^2-33 \zeta (3)\right)    \nonumber\hfill \\ & 
+\frac{1}{3} (x-1) H(0;x) \left(6 \zeta (3) (9-17 x)+\pi ^2 (5 x-3)\right)    \nonumber\hfill \\ & 
+\frac{2}{3} (x-1) H(1;x) \left(6 \zeta (3) (7-4 x)+\pi ^2 (x-1)\right)    \nonumber\hfill \\ & 
-\frac{1}{3} \pi ^2 (x-1) (5 x-3) H(0,1;x)-6 (x-1) (13 x-7) H(0,0,-1,0;x)    \nonumber\hfill \\ & 
+12 (x-1) (5 x-3) H(0,-1,0;x)-6 \pi ^2 (x-1)^2 H(-1,-1;x)    \nonumber\hfill \\ & 
+2 \left(-3+2 \pi ^2\right) H(-1,0;x) (x-1)^2+2 \pi ^2 H(-1,1;x) (x-1)^2    \nonumber\hfill \\ & 
-\frac{1}{6} (x-1) \left(-30 x+\pi ^2 (13 x-7)+18\right) H(0,0;x)    \nonumber\hfill \\ & 
+2 \pi ^2 (x-1)^2 H(1,-1;x)-\frac{1}{3} (x-1) \left(-6 x+\pi ^2 (3 x-5)+6\right) H(1,0;x)    \nonumber\hfill \\ & 
-\frac{2}{3} \pi ^2 H(1,1;x) (x-1)^2-4 (5 x-3) H(0,1,0;x) (x-1)    \nonumber\hfill \\ & 
+2 (x-1) (13 x-7) H(0,0,1,0;x)+36 (x-1) (5 x-3) H(0,-1,-1,0;x)    \nonumber\hfill \\ & 
-24 (x-1) (5 x-3) H(0,-1,0,0;x)-12 (x-1) (5 x-3) H(0,-1,1,0;x)    \nonumber\hfill \\ & 
+144 (x-1)^2 H(-1,0,-1,0;x)-72 (x-1)^2 H(-1,-1,0;x)    \nonumber\hfill \\ & 
+48 H(-1,0,0;x) (x-1)^2+24 H(-1,1,0;x) (x-1)^2    \nonumber\hfill \\ & 
-48 H(-1,0,1,0;x) (x-1)^2-2 (13 x-7) H(0,0,0;x) (x-1)    \nonumber\hfill \\ & 
+24 (x-1)^2 H(1,-1,0;x)-12 (x-1) (3 x-5) H(1,0,-1,0;x)    \nonumber\hfill \\ & 
-8 H(1,1,0;x) (x-1)^2-4 (3 x-5) H(1,0,0;x) (x-1)    \nonumber\hfill \\ & 
+4 (x-1) (3 x-5) H(1,0,1,0;x)-12 (x-1) (5 x-3) H(0,1,-1,0;x)    \nonumber\hfill \\ & 
+2 (x-1) (27 x-17) H(0,1,0,0;x)+4 (x-1) (5 x-3) H(0,1,1,0;x)    \nonumber\hfill \\ & 
+144 (x-1)^2 H(-1,-1,0,0;x)-216 (x-1)^2 H(-1,-1,-1,0;x)    \nonumber\hfill \\ & 
+72 (x-1)^2 H(-1,-1,1,0;x)-60 (x-1)^2 H(-1,0,0,0;x)    \nonumber\hfill \\ & 
+72 (x-1)^2 H(-1,1,-1,0;x)-48 (x-1)^2 H(-1,1,0,0;x)    \nonumber\hfill \\ & 
+(x-1) (29 x-15) H(0,0,0,0;x)-24 (x-1)^2 H(-1,1,1,0;x)    \nonumber\hfill \\ & 
+72 (x-1)^2 H(1,-1,-1,0;x)-48 (x-1)^2 H(1,-1,0,0;x)    \nonumber\hfill \\ & 
+2 (x-1) (7 x-13) H(1,0,0,0;x)-24 (x-1)^2 H(1,-1,1,0;x)    \nonumber\hfill \\ & 
+4 (x-1) (5 x-3) H(1,1,0,0;x)-24 (x-1)^2 H(1,1,-1,0;x)    \nonumber\hfill \\ & 
+8 (x-1)^2 H(1,1,1,0;x)\Big\} \hfill
\end{align}

\subsection{Four propagator integrals}
\begin{align}
{ \SetScale{0.9} \mpfour{20}} &= \int \!\! \frac{d^dk}{i \pi^{d/2}} \int \!\! \frac{d^dl}{i \pi^{d/2}}\frac{1}{D_{11}D_{14}D_{15}D_{17}}\nonumber\\ 
&=\left(\frac{\Gamma(1+\epsilon)}{1-\epsilon}\right)^2 \, (\mtop^2)^{-2 \epsilon }\sum_{i=-2}^{1} \epsilon^{i} F_{3}^{i} (x) + \mathcal{O}(\epsilon^{2})\end{align}
\begin{align}
F_{3}^{-2}(x)\quad = \quad &\frac{1}{2}\hfill\\
F_{3}^{-1}(x)\quad = \quad &-\frac{1}{2}\hfill\\
F_{3}^{0}(x)\quad = \quad &\frac{1}{(1-x)^2}\Big\{-3 x^2+(6-4 \zeta (3)) x+2 \left(x^2-1\right) H(0;x)-3    \nonumber\hfill \\ & 
-H(0,0;x) (x-1)^2+2 x H(0,0,0;x)+4 x H(1,0,0;x)\Big\} \hfill\\
F_{3}^{1}(x)\quad = \quad &\frac{1}{(1-x)^2}\Big\{3 (-4+\zeta (3)) x^2-2 (-12+\zeta (3)) x+\frac{2 \pi ^4 x}{45}-\frac{1}{3} \pi ^2 \left(x^2-1\right)+3 (-4+\zeta (3))    \nonumber\hfill \\ & 
+\frac{1}{6} H(0;x) \left(\pi ^2 (x-1)^2+12 \left(4 x^2-3 \zeta (3) x-4\right)\right)-12 x H(1;x) \zeta (3)    \nonumber\hfill \\ & 
+6 H(0,-1,0;x) (x-1)^2-12 x H(0,0,-1,0;x)-12 \left(x^2-1\right) H(-1,0;x)    \nonumber\hfill \\ & 
+\left(11 x^2-\frac{1}{3} \left(6+\pi ^2\right) x-5\right) H(0,0;x)+\left(4 x^2-\frac{2 \pi ^2 x}{3}-4\right) H(1,0;x)    \nonumber\hfill \\ & 
-2 H(0,1,0;x) (x-1)^2+4 x H(0,0,1,0;x)+\left(-3 x^2+4 x-3\right) H(0,0,0;x)    \nonumber\hfill \\ & 
-24 x H(1,0,-1,0;x)+2 \left(x^2-4 x+1\right) H(1,0,0;x)+8 x H(1,0,1,0;x)    \nonumber\hfill \\ & 
-4 x H(0,1,0,0;x)+6 x H(0,0,0,0;x)+12 x H(1,0,0,0;x)    \nonumber\hfill \\ & 
-8 x H(1,1,0,0;x)\Big\} \hfill
\end{align}

\begin{align}
{ \SetScale{0.9} \tria{20}} &= \int \!\! \frac{d^dk}{i \pi^{d/2}} \int \!\! \frac{d^dl}{i \pi^{d/2}}\frac{1}{D_{12}D_{14}D_{16}D_{17}}\nonumber\\ 
&=\left(\frac{\Gamma(1+\epsilon)}{1-\epsilon}\right)^2 \, (\mtop^2)^{-2 \epsilon }\sum_{i=-2}^{1} \epsilon^{i} F_{4}^{i} (x) + \mathcal{O}(\epsilon^{2})\end{align}
\begin{align}
F_{4}^{-2}(x)\quad = \quad &\frac{1}{2}\hfill\\
F_{4}^{-1}(x)\quad = \quad &\frac{1}{(1-x)^2}\Big\{\frac{3}{2} (x-1)^2+\left(1-x^2\right) H(0;x)\Big\} \hfill\\
F_{4}^{0}(x)\quad = \quad &\frac{1}{(1-x)^2}\Big\{5 x^2+2 (-5+2 \zeta (3)) x+\left(-3 x^2+\frac{\pi ^2 x}{3}+3\right) H(0;x)+5    \nonumber\hfill \\ & 
+2 \left(x^2-1\right) H(-1,0;x)-(x-1) (x+2) H(0,0;x)+\left(1-x^2\right) H(1,0;x)    \nonumber\hfill \\ & 
+2 x H(0,1,0;x)+x H(0,0,0;x)\Big\} \hfill\\
F_{4}^{1}(x)\quad = \quad &\frac{1}{(1-x)^2}\Big\{-(-16+\zeta (3)) x^2+\left(-32-\frac{11 \pi ^4}{90}+\zeta (3)\right) x-4 (-4+\zeta (3))    \nonumber\hfill \\ & 
+H(0;x) \left(-10 x^2-\frac{1}{6} \pi ^2 (x+1) x-3 \zeta (3) x+10\right)-\frac{2}{3} \pi ^2 x H(0,-1;x)    \nonumber\hfill \\ & 
+\frac{1}{3} \pi ^2 x H(0,1;x)+\frac{1}{6} H(1;x) \left(-48 \zeta (3) x-\pi ^2 \left(x^2-1\right)\right)    \nonumber\hfill \\ & 
-2 x H(0,0,-1,0;x)+2 (x-1) (x+2) H(0,-1,0;x)+6 \left(x^2-1\right) H(-1,0;x)    \nonumber\hfill \\ & 
+\left(-3 x^2+\frac{1}{6} \left(-18+\pi ^2\right) x+6\right) H(0,0;x)+\left(-3 x^2-\frac{2 \pi ^2 x}{3}+3\right) H(1,0;x)    \nonumber\hfill \\ & 
-2 \left(x^2+x-1\right) H(0,1,0;x)+2 x H(0,0,1,0;x)-2 x H(0,-1,0,0;x)    \nonumber\hfill \\ & 
-4 x H(0,-1,1,0;x)+\left(4-4 x^2\right) H(-1,-1,0;x)+3 \left(x^2-1\right) H(-1,0,0;x)    \nonumber\hfill \\ & 
+2 \left(x^2-1\right) H(-1,1,0;x)+\left(-2 x^2-3 x+4\right) H(0,0,0;x)    \nonumber\hfill \\ & 
+2 \left(x^2-1\right) H(1,-1,0;x)+\left(-x^2-2 x+3\right) H(1,0,0;x)    \nonumber\hfill \\ & 
+\left(2-2 x^2\right) H(1,1,0;x)-4 x H(1,0,1,0;x)-4 x H(0,1,-1,0;x)    \nonumber\hfill \\ & 
+4 x H(0,1,0,0;x)+4 x H(0,1,1,0;x)+3 x H(0,0,0,0;x)    \nonumber\hfill \\ & 
-2 x H(1,0,0,0;x)\Big\} \hfill
\end{align}

\begin{align}
{ \SetScale{0.9} \triatwo{20}} &= \int \!\! \frac{d^dk}{i \pi^{d/2}} \int \!\! \frac{d^dl}{i \pi^{d/2}}\frac{(k+p_1)\cdot(l-k)}{D_{12}D_{14}D_{16}D_{17}}\nonumber\\ 
&=\left(\frac{\Gamma(1+\epsilon)}{1-\epsilon}\right)^2 \, (\mtop^2)^{1-2 \epsilon }\sum_{i=-2}^{1} \epsilon^{i} F_{5}^{i} (x) + \mathcal{O}(\epsilon^{2})\end{align}
\begin{align}
F_{5}^{-2}(x)\quad = \quad &\frac{(x-1)^2}{8 x}\hfill\\
F_{5}^{-1}(x)\quad = \quad &\frac{1}{(1-x)^2 x}\Big\{\frac{5}{16} (x-1)^4+\left(-\frac{x^4}{4}+x^3-x+\frac{1}{4}\right) H(0;x)\Big\} \hfill\\
F_{5}^{0}(x)\quad = \quad &\frac{1}{(1-x)^2 x}\Big\{\frac{1}{32} \left(31 x^4-140 x^3+(218-64 \zeta (3)) x^2-140 x+31\right)    \nonumber\hfill \\ & 
+\frac{1}{24} \left(-15 x^4+54 x^3-4 \pi ^2 x^2-54 x+15\right) H(0;x)    \nonumber\hfill \\ & 
+\frac{1}{2} \left(x^4-4 x^3+4 x-1\right) H(-1,0;x)    \nonumber\hfill \\ & 
+\frac{1}{4} \left(-x^4+4 x^3+3 x^2-8 x+2\right) H(0,0;x)+\left(-\frac{x^4}{4}+x^3-x+\frac{1}{4}\right) H(1,0;x)    \nonumber\hfill \\ & 
-H(0,1,0;x) x^2-\frac{1}{2} H(0,0,0;x) x^2\Big\} \hfill\\
F_{5}^{1}(x)\quad = \quad &\frac{1}{(1-x)^2 x}\Big\{\left(\frac{189}{64}-\frac{\zeta (3)}{4}\right) x^4+\left(-\frac{233}{16}+\frac{\pi ^2}{24}+\zeta (3)\right) x^3+\left(\frac{743}{32}+\frac{11 \pi ^4}{180}-\frac{3 \zeta (3)}{4}\right) x^2 \nonumber \hfill \\&
-\frac{1}{48} \left(699+2 \pi ^2-192 \zeta (3)\right) x-\zeta (3)+\frac{189}{64}    
+\frac{1}{3} \pi ^2 x^2 H(0,-1;x)    \nonumber\hfill \\ & 
+\frac{1}{48} H(0;x) \left(-\left(93+2 \pi ^2\right) x^4+\left(342+8 \pi ^2\right) x^3+6 \left(\pi ^2+12 \zeta (3)\right) x^2-342 x+93\right)    \nonumber\hfill \\ & 
+\frac{1}{24} H(1;x) \left(96 x^2 \zeta (3)-\pi ^2 \left(x^4-4 x^3+4 x-1\right)\right)    \nonumber\hfill \\ & 
+x^2 H(0,0,-1,0;x)-\frac{1}{6} \pi ^2 x^2 H(0,1;x)    \nonumber\hfill \\ & 
+\frac{1}{2} \left(x^4-4 x^3-3 x^2+8 x-2\right) H(0,-1,0;x)    \nonumber\hfill \\ & 
+\frac{1}{4} \left(5 x^4-18 x^3+18 x-5\right) H(-1,0;x)    \nonumber\hfill \\ & 
+\left(-\frac{5 x^4}{8}+2 x^3-\frac{1}{24} \left(-57+2 \pi ^2\right) x^2-5 x+\frac{5}{4}\right) H(0,0;x)    \nonumber\hfill \\ & 
+\frac{1}{24} \left(-15 x^4+60 x^3+8 \pi ^2 x^2-60 x+15\right) H(1,0;x)    \nonumber\hfill \\ & 
+\frac{1}{2} \left(-x^4+4 x^3+3 x^2-4 x+1\right) H(0,1,0;x)-x^2 H(0,0,1,0;x)    \nonumber\hfill \\ & 
+H(0,-1,0,0;x) x^2+2 H(0,-1,1,0;x) x^2+\left(-x^4+4 x^3-4 x+1\right) H(-1,-1,0;x)    \nonumber\hfill \\ & 
+\frac{3}{4} \left(x^4-4 x^3+4 x-1\right) H(-1,0,0;x)    \nonumber\hfill \\ & 
+\frac{1}{2} \left(x^4-4 x^3+4 x-1\right) H(-1,1,0;x)    \nonumber\hfill \\ & 
+\left(-\frac{x^4}{2}+2 x^3+\frac{9 x^2}{4}-4 x+1\right) H(0,0,0;x)    \nonumber\hfill \\ & 
+\frac{1}{2} \left(x^4-4 x^3+4 x-1\right) H(1,-1,0;x)    \nonumber\hfill \\ & 
+\left(-\frac{x^4}{4}+x^3+\frac{3 x^2}{2}-3 x+\frac{3}{4}\right) H(1,0,0;x)    \nonumber\hfill \\ & 
+2 H(1,0,1,0;x) x^2+\frac{1}{2} \left(-x^4+4 x^3-4 x+1\right) H(1,1,0;x)    \nonumber\hfill \\ & 
+2 H(0,1,-1,0;x) x^2-2 H(0,1,0,0;x) x^2-2 H(0,1,1,0;x) x^2    \nonumber\hfill \\ & 
+x^2 H(1,0,0,0;x)-\frac{3}{2} x^2 H(0,0,0,0;x)\Big\} \hfill
\end{align}

\begin{align}
{ \SetScale{0.9} \triathree{20}} &= \int \!\! \frac{d^dk}{i \pi^{d/2}} \int \!\! \frac{d^dl}{i \pi^{d/2}}\frac{1}{D_{12}D_{14}D_{16}D_{17}^{3}}\nonumber\\ 
&=\left(\frac{\Gamma(1+\epsilon)}{1-\epsilon}\right)^2 \, (\mtop^2)^{-2 \epsilon -2}\sum_{i=0}^{2} \epsilon^{i} F_{6}^{i} (x) + \mathcal{O}(\epsilon^{3})\end{align}
\begin{align}
F_{6}^{0}(x)\quad = \quad &\frac{x H(0,0;x)}{2 (x-1)^2}\hfill\\
F_{6}^{1}(x)\quad = \quad &\frac{x}{(1-x)^2}\Big\{-\frac{1}{4} \pi ^2 H(0;x)-H(0,-1,0;x)-H(0,0;x)-\frac{9 \zeta (3)}{2}    \nonumber\hfill \\ & 
-H(0,1,0;x)+\frac{1}{2} H(0,0,0;x)+H(1,0,0;x)\Big\} \hfill\\
F_{6}^{2}(x)\quad = \quad &\frac{x}{(1-x)^2}\Big\{\frac{1}{2} \pi ^2 H(0,-1;x)+\frac{1}{2} \pi ^2 H(0;x)+9 \zeta (3)+\frac{11 \pi ^4}{144}    \nonumber\hfill \\ & 
+3 \zeta (3) H(1;x)-\frac{1}{6} \pi ^2 H(0,1;x)-H(0,0,-1,0;x)+2 H(0,-1,0;x)    \nonumber\hfill \\ & 
+\frac{1}{12} \left(6-\pi ^2\right) H(0,0;x)+\frac{1}{2} \pi ^2 H(1,0;x)+2 H(0,1,0;x)    \nonumber\hfill \\ & 
+2 H(0,-1,-1,0;x)+2 H(0,-1,1,0;x)-H(0,0,0;x)-2 H(1,0,-1,0;x)    \nonumber\hfill \\ & 
-2 H(1,0,0;x)+4 H(1,0,1,0;x)+2 H(0,1,-1,0;x)-H(0,1,0,0;x)    \nonumber\hfill \\ & 
-2 H(0,1,1,0;x)+\frac{1}{2} H(0,0,0,0;x)+4 H(1,0,0,0;x)+2 H(1,1,0,0;x)\Big\} \hfill
\end{align}

\subsection{Five propagator integrals}
\begin{align}
{ \SetScale{0.9} \mpfive{20}} &= \int \!\! \frac{d^dk}{i \pi^{d/2}} \int \!\! \frac{d^dl}{i \pi^{d/2}}\frac{1}{D_{22}D_{23}D_{24}D_{26}D_{27}}\nonumber\\ 
&=\left(\frac{\Gamma(1+\epsilon)}{1-\epsilon}\right)^2 \, (\mtop^2)^{-2 \epsilon -1} F_{7}^{0} (x) + \mathcal{O}(\epsilon^{1})\end{align}
\begin{align}
F_{7}^{0}(x)\quad = \quad &\frac{1}{(1-x)^2}\Big\{-\frac{1}{6} \pi ^2 H(0,0;x) x-\frac{1}{3} \pi ^2 H(1,0;x) x-\frac{\pi ^4 x}{36}    \nonumber\hfill \\ & 
-x H(0,0,1,0;x)-2 x H(1,0,1,0;x)-2 x H(0,1,0,0;x)-3 x H(1,0,0,0;x)    \nonumber\hfill \\ & 
-4 x H(1,1,0,0;x)\Big\} \hfill
\end{align}

\begin{align}
{ \SetScale{0.9} \dtria{20}} &= \int \!\! \frac{d^dk}{i \pi^{d/2}} \int \!\! \frac{d^dl}{i \pi^{d/2}}\frac{1}{D_{11}D_{13}D_{14}D_{16}D_{17}}\nonumber\\ 
&=\left(\frac{\Gamma(1+\epsilon)}{1-\epsilon}\right)^2 \, (\mtop^2)^{-2 \epsilon -1}\sum_{i=0}^{1} \epsilon^{i} F_{8}^{i} (x) + \mathcal{O}(\epsilon^{2})\end{align}
\begin{align}
F_{8}^{0}(x)\quad = \quad &\frac{x}{(1-x)^2}\Big\{-2 H(0,0,1;x)-2 H(0,1,0;x)+4 H(1,0,0;x)-6 \zeta (3)\Big\} \hfill\\
F_{8}^{1}(x)\quad = \quad &\frac{x}{(1-x)^2}\Big\{-12 \zeta (3) H(0;x)+\frac{1}{3} \pi ^2 H(0,1;x)-24 H(1;x) \zeta (3)-\frac{\pi ^4}{10}    \nonumber\hfill \\ & 
-8 H(0,0,0,1;x)-10 H(0,0,-1,0;x)+4 H(0,-1,0,1;x)-\frac{2}{3} \pi ^2 H(1,0;x)    \nonumber\hfill \\ & 
-4 H(1,0,0,1;x)-4 H(0,1,0,1;x)-4 H(0,0,1,0;x)-4 H(0,0,1,1;x)+4 H(0,-1,0,0;x)    \nonumber\hfill \\ & 
+4 H(0,-1,1,0;x)-24 H(1,0,-1,0;x)+4 H(1,0,1,0;x)+4 H(0,1,-1,0;x)    \nonumber\hfill \\ & 
-6 H(0,1,0,0;x)-4 H(0,1,1,0;x)+12 H(1,0,0,0;x)\Big\} \hfill
\end{align}

\subsection{Six propagator integrals}
\begin{align}
{ \SetScale{0.9} \mpsix{20}} &= \int \!\! \frac{d^dk}{i \pi^{d/2}} \int \!\! \frac{d^dl}{i \pi^{d/2}}\frac{1}{D_{21}D_{23}D_{24}D_{25}D_{26}D_{27}}\nonumber\\ 
&=\left(\frac{\Gamma(1+\epsilon)}{1-\epsilon}\right)^2 \, (\mtop^2)^{-2 \epsilon -2} F_{9}^{0} (x) + \mathcal{O}(\epsilon^{1})\end{align}
\begin{align}
F_{9}^{0}(x)\quad = \quad &\frac{x^2}{(1-x)^3 (x+1)}\Big\{8 \zeta (3) H(0;x)+16 H(0,0,-1,0;x)+\frac{\pi ^4}{10}    \nonumber\hfill \\ & 
+\frac{2}{3} \pi ^2 H(0,0;x)-4 H(0,0,1,0;x)-8 H(0,-1,0,0;x)+14 H(0,1,0,0;x)    \nonumber\hfill \\ & 
+H(0,0,0,0;x)\Big\} \hfill
\end{align}

\begin{align}
{ \SetScale{0.9} \xtria{20}} &= \int \!\! \frac{d^dk}{i \pi^{d/2}} \int \!\! \frac{d^dl}{i \pi^{d/2}}\frac{1}{D_{31}D_{32}D_{33}D_{34}D_{35}D_{37}}\nonumber\\ 
&=\left(\frac{\Gamma(1+\epsilon)}{1-\epsilon}\right)^2 \, (\mtop^2)^{-2 \epsilon -2}\sum_{i=-1}^{0} \epsilon^{i} F_{10}^{i} (x) + \mathcal{O}(\epsilon^{1})\end{align}
\begin{align}
F_{10}^{-1}(x)\quad = \quad &\frac{x^2}{(1-x)^4}\Big\{-\frac{2}{3} \pi ^2 H(0;x)-8 H(0,-1,0;x)+4 H(0,0,0;x)-12 \zeta (3)\Big\} \hfill\\
F_{10}^{0}(x)\quad = \quad &\frac{x^2}{(1-x)^4}\Big\{\frac{8}{3} \pi ^2 H(0,-1;x)+24 \zeta (3)-\frac{16 \pi ^4}{45}    \nonumber\hfill \\ & 
+\frac{4}{3} \left(\pi ^2-33 \zeta (3)\right) H(0;x)-\frac{4}{3} \pi ^2 H(0,1;x)-48 H(1;x) \zeta (3)    \nonumber\hfill \\ & 
-56 H(0,0,-1,0;x)+16 H(0,-1,0;x)-\frac{10}{3} \pi ^2 H(0,0;x)-\frac{8}{3} \pi ^2 H(1,0;x)    \nonumber\hfill \\ & 
+8 H(0,0,1,0;x)+64 H(0,-1,-1,0;x)-40 H(0,-1,0,0;x)-16 H(0,-1,1,0;x)    \nonumber\hfill \\ & 
-8 H(0,0,0;x)-32 H(1,0,-1,0;x)-16 H(0,1,-1,0;x)+8 H(0,1,0,0;x)    \nonumber\hfill \\ & 
+12 H(0,0,0,0;x)+16 H(1,0,0,0;x)\Big\} \hfill
\end{align}

\section{Analytic continuation}
\label{sec:2analytic}
The expressions in section \ref{sec:1result} correspond to the unphysical,
space-like region. The results must be analytically continued towards
the time-like region. Due to the threshold in $s=4m_t^2$, the physical 
region splits into two subregions, namely above and below threshold.

\subsection{Analytic continuation above threshold}

The region above threshold corresponds to the range $-1<x<0$.
The analytic continuation to this region is straightforward, using 
the properties of harmonic polylogarithms under the transformation $x \to
-x$ \cite{Remiddi:1999ew,analcont1}. 
For harmonic polylogarithms $H(a_n,\ldots,a_1;x+i \vep)$ with $a_1 \neq 0$ the 
analytic continuation is obtained trivially  
\beq
H(a_n,\ldots,\pm 1 ;x+i \vep)  =
(-1)^{\pm 1+\cdots+a_n}H(-a_n,\ldots,\mp 1;-x) 
\label{eq:xtominusx2}
\ .
\eeq
where $a_k = -1,0,1$ for $k\neq 1$.
We can eliminate higher rank harmonic polylogarithms with  
$a_1 = 0 $ by applying integration by parts and product identities \cite{Remiddi:1999ew}.    
For instance:
\beq
H(1,0;x) = H(0;x)\, H(1;x) - H(0,1;x) \ . 
\eeq
Recursively, one can write similar identities for the harmonic 
polylogarithms of higher rank. At the end of this procedure 
we only require the analytic continuation of simple logarithms: 
\bea
H(1;x+i\vep) & = & - H(-1;-x+i\vep) \nonumber \\
H(0;x+i\vep) & = & H(0;-x+i\vep) +  i\pi\nonumber  \\
H(-1;x+i\vep)& = & - H(1;-x+i\vep)  \ . 
\label{eq:xtominusx1}
\eea
The analytic continuation described here is incorporated 
in the Mathematica package~\cite{Maitre:2005uu}. The 
expressions for the master integrals above threshold can be easily 
obtained from the ones in the space-like region using the routines
implemented in this package.

\subsection{Analytic continuation below threshold}

Below threshold, the variable $x$ lies on the unit circle of the complex
plane. For this analytic continuation we follow the  procedure  
in~\cite{Davydychev:2003mv}, which we summarize here.

We first express our results in terms of the variable $\theta$ given by
$x=\exp(i\,\theta)$, and introduce the following notation for the harmonic
polylogarithms as functions of $\theta$:
\beq
H_c(a_n,\ldots,a_1;\theta) \stackrel{\mathrm{def}}{=} H(a_n,\ldots,a_1;e^{i\theta}) \ .
\eeq
We now eliminate 
polylogarithms  $H_c(a_n,\ldots,a_1;\theta)$ with $a_n=1$, using integration
by parts and product identities~\cite{analcont1}. 
Then we use the analytic continuation 
of harmonic polylogarithms of weight one  as kernels.  
For $0<\theta<\pi$ we have
\bea
H_c(1;\theta) & = & -\ln 2\left|\sin\frac{\theta}{2}\right| \ + \ i\left(\frac{\pi}{2} -
  \frac{\theta}{2}\right) \label{eq:HPLc1} \\
H_c(0;\theta) & = & i \theta \label{eq:HPLc0} \\
H_c(-1;\theta) & = & \ln 2\left|\cos\frac{\theta}{2}\right| + i \frac{\theta}{2} \label{eq:HPLc-1}
\eea
We can find the analytic continuation for the 
harmonic polylogarithms of higher weights recursively, using

\beq
H_c(a_n,a_{n-1},\ldots,a_1;\theta) = H(a_n,a_{n-1},\ldots,a_1;1) + i \int_0^{\theta}
\ud {\theta}^{\prime} g(a_n;{\theta}^{\prime}) H_c(a_{n-1},\ldots,a_1;{\theta}^{\prime}) \label{eq:hc_def}
\eeq 
where
\bea
g(1;\theta) & = &  \frac{e^{i\theta}}{1-e^{i\theta}} =
-\frac{1}{2}\  + \ i\frac{1}{2}\cot\frac{\theta}{2} \ ,\\
g(0;\theta) & = & 
\frac{e^{i\theta}}{e^{i\theta}} = 1 \ ,\\
g(-1;\theta) & = & \frac{e^{i\theta}}{1+e^{i\theta}} =
\frac{1}{2}\ + \   i\frac{1}{2}\tan\frac{\theta}{2} \ .
\eea
In this way, we have obtained expressions for the analytically continued 
harmonic polylogarithms through weight 4. 
The analytically continued master integrals can be expressed in terms of 
the  following functions \cite{Kalmykov:2004xg,Kalmykov:2005hb,Davydychev:2003mv}:
\bea
\Cl(1,\theta) & = & -\ln 2\left|\sin\frac{\theta}{2}\right| \ , \\
\LS(j,k,\theta) & = & - \int_0^{\theta} \ud {{\theta}^{\prime}} \, {{\theta}^{\prime}}^k \ln^{j-k-1}
2\left|\sin\frac{{\theta}^{\prime}}{2}\right| \ , \\
\LSC(i,j,\theta) & = & - \int_0^{\theta} \ud {{\theta}^{\prime}} \, \ln^{i-1}
2\left|\sin\frac{{{\theta}^{\prime}}}{2}\right| \ln^{j-1}
2\left|\cos\frac{{{\theta}^{\prime}}}{2}\right| \ , \\
\LSLSC(n,i,j,\theta) & = & \int_0^{\theta} \ud {{\theta}^{\prime}} \, \LS(n+1,0,{{\theta}^{\prime}})\ln^{i-1} 2\left|\sin\frac{{{\theta}^{\prime}}}{2}\right| \ln^{j-1}
2\left|\cos\frac{{{\theta}^{\prime}}}{2}\right| \ .
\eea
Interestingly, these functions are a smaller set than the functions 
that appear  in the analytic continuation of individual harmonic 
polylogarithms.

\section{Results for the master integrals in the region below threshold}
\label{sec:MIcircle}
\allowdisplaybreaks[2] 
We now present the master integrals in the kinematic region $0 < s < 4 m_t^2$. 
The expressions $\tilde F_j^i(\theta)$ correspond to the analytic continuation 
of  the coefficients  $F_j^i(x)$ in section~\ref{sec:1result}.

\subsection{One loop integrals}
\begin{align}
\tilde F_{\mathrm{mbub}}^{-1}(\theta)\quad = \quad &1\\
\tilde F_{\mathrm{mbub}}^{0}(\theta)\quad = \quad &-\theta  \cot \left(\frac{\theta }{2}\right)+1\\
\tilde F_{\mathrm{mbub}}^{1}(\theta)\quad = \quad &\cot \left(\frac{\theta }{2}\right)\Big(-\theta -2 \theta  \Cl(1,\theta -\pi )+2 \LS(2,0,\theta -\pi )\Big)+2\\
\tilde F_{\mathrm{mbub}}^{2}(\theta)\quad = \quad &\frac{1}{6} \cot \left(\frac{\theta }{2}\right)\Big(6 \Big(-2 \theta -\frac{\pi ^3}{6}\Big)-12 \theta  \Cl(1,\theta -\pi )-12 \theta  \Cl(1,\theta -\pi )^2    \nonumber\hfill \\ & 
+(24 \Cl(1,\theta -\pi )+12) \LS(2,0,\theta -\pi )+12 \LS(3,0,\theta -\pi )\Big)+4\\
\tilde F_{\mathrm{mbub}}^{3}(\theta)\quad = \quad &\frac{1}{6} \cot \left(\frac{\theta }{2}\right)\Big(-\pi ^3-24 \theta +6 \Big(-4 \theta -\frac{\pi ^3}{3}\Big) \Cl(1,\theta -\pi )-12 \theta  \Cl(1,\theta -\pi )^2    \nonumber\hfill \\ & 
-8 \theta  \Cl(1,\theta -\pi )^3+\left(24 \Cl(1,\theta -\pi )^2+24 \Cl(1,\theta -\pi )+24\right) \LS(2,0,\theta -\pi )    \nonumber\hfill \\ & 
+(24 \Cl(1,\theta -\pi )+12) \LS(3,0,\theta -\pi )+8 \LS(4,0,\theta -\pi )+12 \pi  \zeta (3)\Big)+8
\end{align}

\begin{align}
\tilde F_{\mathrm{mtri}}^{0}(\theta)\quad = \quad &-\frac{1}{8} \theta ^2 \csc ^2\left(\frac{\theta }{2}\right)\\
\tilde F_{\mathrm{mtri}}^{1}(\theta)\quad = \quad &\frac{1}{8} \csc ^2\left(\frac{\theta }{2}\right)\Big(\theta ^2+(4 \theta -8 \pi ) \LS(2,0,\theta -\pi )-8 \LS(3,1,\theta -\pi )-14 \zeta (3)\Big)\\
\tilde F_{\mathrm{mtri}}^{2}(\theta)\quad = \quad &\frac{1}{24} \csc ^2\left(\frac{\theta }{2}\right)\Big(-\pi ^3 \theta +12 (2 \pi -\theta ) \LS(2,0,\theta -\pi )+12 \LS(2,0,\theta -\pi )^2    \nonumber\hfill \\ & 
+12 \theta  \LS(3,0,\theta -\pi )+24 \LS(3,1,\theta -\pi )+42 \zeta (3)\Big)
\end{align}

\subsection{Three propagator integrals}
\begin{align}
\tilde F_{1}^{0}(\theta)\quad = \quad &-\frac{1}{4} \theta ^2 \csc ^2\left(\frac{\theta }{2}\right)\hfill\\
\tilde F_{1}^{1}(\theta)\quad = \quad &\frac{1}{2} \csc ^2\left(\frac{\theta }{2}\right)\Big(\theta ^2+\theta ^2 \Cl(1,\theta )+2 \theta  \LS(2,0,\theta )+6 \LS(2,0,\theta -\pi ) 
(-2 \pi +\theta )   \nonumber\hfill \\ & 
-6 \LS(3,1,\theta )-12 \LS(3,1,\theta -\pi )-21 \zeta (3)\Big) \hfill\\
\tilde F_{1}^{2}(\theta)\quad = \quad &\frac{1}{240} \csc ^2\left(\frac{\theta }{2}\right)\Big(79 \pi ^4-180 \pi ^3 \theta -60 \theta ^2-30 \theta ^4-120\theta ^2 \Cl(1,\theta )^2   +120 \pi ^2 \log ^2(2)   -120 \log ^4(2)
\nonumber\hfill \\ & 
-480 \theta  \LS(2,0,\theta )+480 \LS(2,0,\theta )^2+\LS(2,0,\theta -\pi)\big(2880 \pi -1440 \theta  +2880 \LS(2,0,\theta )\big) \nonumber\hfill \\ & 
+2160 \LS(2,0,\theta-\pi )^2  +240 \theta  \LS(3,0,\theta )+720 \LS(3,0,\theta -\pi )\big(-\pi +3
\theta \big)   +1440 \LS(3,1,\theta ) \nonumber\hfill \\ & 
+2880 \LS(3,1,\theta -\pi )+180 \LS(4,1,2 \theta )-720 \LS(4,1,\theta -\pi)+1440 \theta  \LSC(2,2,\theta )  +1440 \LSLSC(1,1,2,\theta )  \nonumber\hfill
\\ & 
-2880 \text{Li}_4\left(\frac{1}{2}\right)+5040 \zeta (3)-2520 \log (2) \zeta(3)+\Cl(1,\theta )\Big(-240 \theta ^2   -480 \theta  \LS(2,0,\theta )
\nonumber\hfill \\ & 
-1440 (\theta -2 \pi ) \LS(2,0,\theta -\pi )+1440 \LS(3,1,\theta )+2880 \LS(3,1,\theta -\pi )   
+5040 \zeta (3)\Big)\Big) \hfill
\end{align}

\begin{align}
\tilde F_{2}^{-1}(\theta)\quad = \quad &\frac{1}{2} \theta  \csc (\theta )\\
\tilde F_{2}^{0}(\theta)\quad = \quad &\frac{1}{4} \theta ^2 \cot \left(\frac{\theta }{2}\right) \csc (\theta )+\csc (\theta )\Big(-\theta +\theta  \Cl(1,\theta )+3 \theta  \Cl(1,\theta -\pi )   
-\LS(2,0,\theta )-3 \LS(2,0,\theta -\pi )\Big)\\
\tilde F_{2}^{1}(\theta)\quad = \quad &+\frac{1}{2} \cot \left(\frac{\theta  }{2}\right) \csc (\theta )\Big(-\theta ^2-\theta ^2 \Cl(1,\theta )-2 \theta \LS(2,0,\theta )   +(12 \pi -6 \theta ) \LS(2,0,\theta -\pi )+6 \LS(3,1,\theta  ) \nonumber\hfill \\ &
+12 \LS(3,1,\theta -\pi )+21 \zeta (3)\Big)+\frac{\csc (\theta )}{4}\Big(
+3\pi ^3+2 \theta +2 \theta ^3+4 \theta  \Cl(1,\theta )^2-24 \theta \Cl(1,\theta -\pi )  \nonumber\hfill \\ & 
  +36 \theta  \Cl(1,\theta -\pi )^2   +\Cl(1,\theta ) (24 \theta  \Cl(1,\theta -\pi )-8 \theta )+(-8 \Cl(1,\theta )-24 \Cl(1,\theta -\pi )+8) \LS(2,0,\theta )    \nonumber\hfill \\ & 
+(-24 \Cl(1,\theta )-72 \Cl(1,\theta -\pi )+24) \LS(2,0,\theta -\pi )-4 \LS(3,0,\theta )-36 \LS(3,0,\theta -\pi )    \nonumber\hfill \\ & 
-24 \LSC(2,2,\theta )\Big)\\
\tilde F_{2}^{2}(\theta)\quad = \quad &\frac{1}{240} \cot \left(\frac{\theta }{2}\right) \csc (\theta )\Big(-79 \pi ^4+180 \pi ^3 \theta +60 \theta ^2+30 \theta ^4+120 \theta ^2 \Cl(1,\theta )^2    \nonumber\hfill \\ & 
-120 \pi ^2 \log ^2(2)+120 \log ^4(2)-480 \LS(2,0,\theta )^2+\LS(2,0,\theta )
(480 \Cl(1,\theta ) \theta +480 \theta  \nonumber\hfill \\ & 
-2880 \LS(2,0,\theta -\pi ))+(240 (6 \theta -12 \pi )+1440 (\theta -2 \pi ) \Cl(1,\theta )) \LS(2,0,\theta -\pi )-2160 \LS(2,0,\theta -\pi )^2    \nonumber\hfill \\ & 
-240 \theta  \LS(3,0,\theta )+240 (3 \pi -9 \theta ) \LS(3,0,\theta -\pi )+(-1440 \Cl(1,\theta )-1440) \LS(3,1,\theta )    \nonumber\hfill \\ & 
+(-2880 \Cl(1,\theta )-2880) \LS(3,1,\theta -\pi )-180 \LS(4,1,2 \theta )+720 \LS(4,1,\theta -\pi )-1440 \theta  \LSC(2,2,\theta )    \nonumber\hfill \\ & 
-1440 \LSLSC(1,1,2,\theta )+2880 \text{Li}_4\left(\frac{1}{2}\right)+240 \Cl(1,\theta ) \left(\theta ^2-21 \zeta (3)\right)    \nonumber\hfill \\ & 
-5040 \zeta (3)+2520 \log (2) \zeta (3)\Big)+\frac{\csc (\theta )}{6}\Big(4 \theta  \Cl(1,\theta )^3+\left(18 \theta ^3+18 \theta +27 \pi ^3\right) \Cl(1,\theta -\pi )    \nonumber\hfill \\ & 
-108 \theta  \Cl(1,\theta -\pi )^2+108 \theta  \Cl(1,\theta -\pi )^3+\Cl(1,\theta )^2 (36 \theta  \Cl(1,\theta -\pi )-12 \theta )    \nonumber\hfill \\ & 
+\Cl(1,\theta ) \left(6 \theta ^3+108 \Cl(1,\theta -\pi )^2 \theta  -72
  \Cl(1,\theta -\pi ) \theta +6 \theta +9 \pi ^3\right)+\big(-18 \theta ^2-12
  \Cl(1,\theta )^2   \nonumber\hfill \\ & 
-108 \Cl(1,\theta -\pi )^2+\Cl(1,\theta ) (24-72 \Cl(1,\theta -\pi ))+72 \Cl(1,\theta -\pi )-6\big) \LS(2,0,\theta )  \nonumber\hfill \\ & 
+\big(-54 \theta ^2+216 \pi  \theta -36 \Cl(1,\theta )^2-324 \Cl(1,\theta  -\pi )^2+\Cl(1,\theta ) (72-216 \Cl(1,\theta -\pi ))  \nonumber\hfill \\ & 
+216 \Cl(1,\theta -\pi )-216 \pi ^2-18\big) \LS(2,0,\theta -\pi )    +(-12
\Cl(1,\theta )-36 \Cl(1,\theta -\pi )\nonumber\hfill \\ & 
+12) \LS(3,0,\theta )+(-108 \Cl(1,\theta )-324 \Cl(1,\theta -\pi )+108) \LS(3,0,\theta -\pi )    \nonumber\hfill \\ & 
+108 \theta  \LS(3,1,\theta )+(216 \theta -432 \pi ) \LS(3,1,\theta -\pi )-4 \LS(4,0,\theta )-108 \LS(4,0,\theta -\pi )    \nonumber\hfill \\ & 
-126 \LS(4,2,\theta )-216 \LS(4,2,\theta -\pi )+(-72 \Cl(1,\theta )-216 \Cl(1,\theta -\pi )+72) \LSC(2,2,\theta )    \nonumber\hfill \\ & 
-108 \LSC(2,3,\theta )-36 \LSC(3,2,\theta )-3 \left(2 \theta ^3-126 \zeta (3) \theta +198 \pi  \zeta (3)+3 \pi ^3\right)\Big)
\end{align}

\subsection{Four propagator integrals}
\begin{align}
\tilde F_{3}^{-2}(\theta)\quad = \quad &\frac{1}{2}\\
\tilde F_{3}^{-1}(\theta)\quad = \quad &-\frac{1}{2}\\
\tilde F_{3}^{0}(\theta)\quad = \quad &\frac{1}{4} \csc ^2\left(\frac{\theta }{2}\right)\Big(-6+\theta ^2+2 \theta ^2 \Cl(1,\theta )+6 \cos (\theta )-\theta ^2 \cos (\theta )-4 \LS(3,1,\theta )+4 \theta  \sin (\theta ) \Big)\\
\tilde F_{3}^{1}(\theta)\quad = \quad &\frac{1}{240} \csc ^2\left(\frac{\theta }{2}\right)\Big(-1440+79 \pi ^4-60 \theta ^2-120 \theta ^2 \Cl(1,\theta )^2+1440 \cos (\theta )+180 \theta ^2 \cos (\theta )    \nonumber\hfill \\ & 
+120 \pi ^2 \log ^2(2)-120 \log ^4(2)+240 \LS(2,0,\theta )^2-720 \pi \LS(3,0,\theta -\pi )+(1440 \Cl(1,\theta )  \nonumber\hfill \\ &    -240 (3 \cos (\theta )-4)) \LS(3,1,\theta )+(2880 \Cl(1,\theta )-1440 (\cos (\theta )-1)) \LS(3,1,\theta -\pi )+180 \LS(4,1,2 \theta )    \nonumber\hfill \\ & 
-720 \LS(4,1,\theta -\pi )+1440 \LSLSC(1,1,2,\theta )   -2880
\text{Li}_4\left(\frac{1}{2}\right)+\LS(2,0,\theta ) (-480 \theta \Cl(1,\theta ) \nonumber\hfill \\ & 
+1440 \LS(2,0,\theta -\pi )+240 (\cos (\theta ) \theta -\theta -2 \sin (\theta)))  +960 \theta  \sin (\theta )+1440 \theta  \Cl(1,\theta -\pi ) \sin (\theta )  \nonumber\hfill \\ & 
+\LS(2,0,\theta -\pi ) (1440 (2 \pi -\theta ) \Cl(1,\theta )-720 (-\cos (\theta ) \theta +\theta +2 \pi  \cos (\theta )+2 \sin (\theta )-2 \pi ))    \nonumber\hfill \\ & 
+2520 \zeta (3)-2520 \cos (\theta ) \zeta (3)-2520 \log (2) \zeta (3) \nonumber\hfill \\ & 
+120 \Cl(1,\theta ) \left(\cos (\theta ) \theta ^2-2 \theta ^2+4 \sin (\theta ) \theta +42 \zeta (3)\right)  
\Big)
\end{align}

\begin{align}
\tilde F_{4}^{-2}(\theta)\quad = \quad &\frac{1}{2}\\
\tilde F_{4}^{-1}(\theta)\quad = \quad &-\theta  \cot \left(\frac{\theta }{2}\right)+\frac{3}{2}\\
\tilde F_{4}^{0}(\theta)\quad = \quad &\frac{1}{8} \csc ^2\left(\frac{\theta }{2}\right)\Big(20-\theta ^2-20 \cos (\theta )+\theta ^2 \cos (\theta )+8 \LS(3,1,\theta )-4 \LS(2,0,\theta ) (\theta -\sin (\theta ))    \nonumber\hfill \\ & 
-12 \theta  \sin (\theta )-4 \theta  \Cl(1,\theta ) \sin (\theta )-8 \theta  \Cl(1,\theta -\pi ) \sin (\theta )+8 \LS(2,0,\theta -\pi ) \sin (\theta )\Big)\\
\tilde F_{4}^{1}(\theta)\quad = \quad &\frac{1}{96} \csc ^2\left(\frac{\theta }{2}\right)\Big(768-36 \theta ^2-\theta ^4-768 \cos (\theta )+36 \theta ^2 \cos (\theta )-96 \LS(2,0,\theta )^2 +(-192 \Cl(1,\theta )   \nonumber\hfill
\\ & 
-48 (\cos (\theta )+1)) \LS(3,1,\theta )+96 (\cos (\theta )-1) \LS(3,1,\theta -\pi )-96 \LS(4,1,\theta )-48 \LS(3,0,\theta ) (\theta -\sin (\theta ))    \nonumber\hfill \\ & 
-96 \LSC(2,2,\theta ) (\theta -\sin (\theta ))-8 \pi ^3 \sin (\theta )-480 \theta  \sin (\theta )+4 \theta ^3 \sin (\theta )-48 \theta  \Cl(1,\theta )^2 \sin (\theta )    \nonumber\hfill \\ & 
-288 \theta  \Cl(1,\theta -\pi ) \sin (\theta )-96 \theta  \Cl(1,\theta -\pi )^2 \sin (\theta )+96 \LS(3,0,\theta -\pi ) \sin (\theta )    \nonumber\hfill \\ & 
+\LS(2,0,\theta ) (-96 \LS(2,0,\theta -\pi )+96 \Cl(1,\theta -\pi ) \sin (\theta )+96 \Cl(1,\theta ) (\theta +\sin (\theta ))+48 (\theta +3 \sin (\theta )))    \nonumber\hfill \\ & 
+\LS(2,0,\theta -\pi ) (96 \Cl(1,\theta ) \sin (\theta )+192 \Cl(1,\theta -\pi) \sin (\theta )+48 (-\cos (\theta ) \theta +\theta +2 \pi  \cos (\theta )
\nonumber\hfill \\ & 
+6 \sin (\theta )-2 \pi ))  +\Cl(1,\theta ) \left(24 \left(\cos (\theta ) \theta ^2-\theta ^2-6 \sin (\theta ) \theta \right)-96 \theta  \Cl(1,\theta -\pi ) \sin (\theta )\right)-168 \zeta (3)    \nonumber\hfill \\ & 
+168 \cos (\theta ) \zeta (3)\Big)
\end{align}

\begin{align}
\tilde F_{5}^{-2}(\theta)\quad = \quad &-\frac{1}{2} \sin ^2\left(\frac{\theta }{2}\right)\\
\tilde F_{5}^{-1}(\theta)\quad = \quad &\frac{1}{32} \csc ^2\left(\frac{\theta }{2}\right) (20 \cos (\theta )-5 \cos (2 \theta )+16 \theta  \sin (\theta )-4 \theta  \sin (2 \theta )-15)\\
\tilde F_{5}^{0}(\theta)\quad = \quad &\frac{1}{64} \csc ^2\left(\frac{\theta }{2}\right)\Big(-109+6 \theta ^2+140 \cos (\theta )-8 \theta ^2 \cos (\theta )-31 \cos (2 \theta )+2 \theta ^2 \cos (2 \theta )    \nonumber\hfill \\ & 
-32 \LS(3,1,\theta )+72 \theta  \sin (\theta )-16 \LS(2,0,\theta -\pi ) (4 \sin (\theta )-\sin (2 \theta ))-20 \theta  \sin (2 \theta )    \nonumber\hfill \\ & 
+8 \LS(2,0,\theta ) (2 \theta -4 \sin (\theta )+\sin (2 \theta ))+8 \Cl(1,\theta ) (4 \theta  \sin (\theta )-\theta  \sin (2 \theta ))    \nonumber\hfill \\ & 
+16 \Cl(1,\theta -\pi ) (4 \theta  \sin (\theta )-\theta  \sin (2 \theta ))\Big)\\
\tilde F_{5}^{1}(\theta)\quad = \quad &\frac{1}{384} \csc ^2\left(\frac{\theta }{2}\right)\Big(-2229+114 \theta ^2+2 \theta ^4+2796 \cos (\theta )-144 \theta ^2 \cos (\theta )-567 \cos (2 \theta )    \nonumber\hfill \\ & 
+30 \theta ^2 \cos (2 \theta )+192 \LS(2,0,\theta )^2+(384 \Cl(1,\theta )+48 (4 \cos (\theta )-\cos (2 \theta )+3)) \LS(3,1,\theta )    \nonumber\hfill \\ & 
-96 (4 \cos (\theta )-\cos (2 \theta )-3) \LS(3,1,\theta -\pi )+192 \LS(4,1,\theta )+32 \pi ^3 \sin (\theta )+1368 \theta  \sin (\theta )    \nonumber\hfill \\ & 
-16 \theta ^3 \sin (\theta )  +\LS(2,0,\theta -\pi ) (-48 (-4 \cos (\theta )
\theta +\cos (2 \theta ) \theta +3 \theta +8 \pi  \cos (\theta )-2 \pi  \cos (2 \theta )  \nonumber\hfill \\ & 
+18 \sin (\theta )-5 \sin (2 \theta )-6 \pi )-96 \Cl(1,\theta ) (4 \sin (\theta )-\sin (2 \theta )) -192 \Cl(1,\theta -\pi ) (4 \sin (\theta ) \nonumber\hfill \\ & 
-\sin (2 \theta ))) +\LS(2,0,\theta ) (192 \LS(2,0,\theta -\pi )   -24 (6 \theta +20 \sin (\theta )-5 \sin (2 \theta ))\nonumber\hfill \\ & 
-96 \Cl(1,\theta -\pi ) (4 \sin (\theta )-\sin (2 \theta ))-96 \Cl(1,\theta ) (2 \theta +4 \sin (\theta )-\sin (2 \theta )))    \nonumber\hfill \\ & 
-96 \LS(3,0,\theta -\pi ) (4 \sin (\theta )-\sin (2 \theta ))-8 \pi ^3 \sin (2 \theta )-372 \theta  \sin (2 \theta )+4 \theta ^3 \sin (2 \theta )    \nonumber\hfill \\ & 
+48 \LS(3,0,\theta ) (2 \theta -4 \sin (\theta )+\sin (2 \theta ))+96 \LSC(2,2,\theta ) (2 \theta -4 \sin (\theta )+\sin (2 \theta ))    \nonumber\hfill \\ & 
+48 \Cl(1,\theta -\pi ) (18 \theta  \sin (\theta )-5 \theta  \sin (2 \theta))+48 \Cl(1,\theta )^2 (4 \theta  \sin (\theta )-\theta  \sin (2 \theta )) \nonumber\hfill \\ & 
 +96 \Cl(1,\theta -\pi )^2 (4 \theta  \sin (\theta )-\theta  \sin (2 \theta )) +\Cl(1,\theta ) \left(96 \Cl(1,\theta -\pi ) (4 \theta  \sin (\theta )-\theta   \sin (2 \theta )) \right. \nonumber\hfill \\ & 
\left.
-24 \left(4 \cos (\theta ) \theta ^2-\cos (2 \theta ) \theta ^2-3 \theta ^2-20 \sin (\theta ) \theta +5 \sin (2 \theta ) \theta \right)\right)    \nonumber\hfill \\ & 
+504 \zeta (3)-672 \cos (\theta ) \zeta (3)+168 \cos (2 \theta ) \zeta (3)
\Big)
\end{align}

\begin{align}
\tilde F_{6}^{0}(\theta)\quad = \quad &\frac{1}{16} \theta ^2 \csc ^2\left(\frac{\theta }{2}\right)\\
\tilde F_{6}^{1}(\theta)\quad = \quad &\frac{1}{8} \csc ^2\left(\frac{\theta }{2}\right)\Big(-\theta ^2+\theta ^2 \Cl(1,\theta )+2 \theta  \LS(2,0,\theta )+2 (2 \pi -\theta ) \LS(2,0,\theta -\pi )    \nonumber\hfill \\ & 
-6 \LS(3,1,\theta )+4 \LS(3,1,\theta -\pi )+7 \zeta (3)\Big)\\
\tilde F_{6}^{2}(\theta)\quad = \quad &\frac{1}{2880} \csc ^2\left(\frac{\theta }{2}\right) \Big(79 \pi ^4+60 \pi ^3 \theta +180 \theta ^2+15 \theta
^4+360 \theta ^2 \Cl(1,\theta )^2 +120 \pi ^2 \log ^2(2)-120 \log ^4(2)
\nonumber\hfill \\ & 
+2160 \LS(2,0,\theta )^2+(1440 \theta -720 (2 \theta -4 \pi ) \Cl(1,\theta )-2880 \pi ) \LS(2,0,\theta -\pi )    \nonumber\hfill \\ & 
-720 \LS(2,0,\theta -\pi )^2+\LS(2,0,\theta ) (-2880 \Cl(1,\theta ) \theta -1440 \theta +2880 \LS(2,0,\theta -\pi ))+720 \theta  \LS(3,0,\theta )    \nonumber\hfill \\ & 
-720 (\theta +\pi ) \LS(3,0,\theta -\pi )+(4320 \Cl(1,\theta )+4320) \LS(3,1,\theta )+(2880 \Cl(1,\theta )-2880) \LS(3,1,\theta -\pi )    \nonumber\hfill \\ & 
+1440 \LS(4,1,\theta )+180 \LS(4,1,2 \theta )-720 \LS(4,1,\theta -\pi )+1440 \theta  \LSC(2,2,\theta )+1440 \LSLSC(1,1,2,\theta )    \nonumber\hfill \\ & 
-2880 \text{Li}_4\left(\frac{1}{2}\right)-720 \Cl(1,\theta ) \left(\theta ^2-7 \zeta (3)\right)-5040 \zeta (3)-2520 \log (2) \zeta (3)\Big)
\end{align}

\subsection{Five propagator integrals}
\begin{align}
\tilde F_{7}^{0}(\theta)\quad = \quad &\frac{1}{64} \csc ^2\left(\frac{\theta }{2}\right)\Big(-\theta ^4-16 \theta ^2 \Cl(1,\theta )^2+32 \theta  \Cl(1,\theta ) \LS(2,0,\theta )-16 \LS(2,0,\theta )^2    \Big) \hfill
\end{align}

\begin{align}
\tilde F_{8}^{0}(\theta)\quad = \quad &\frac{1}{8} \csc ^2\left(\frac{\theta }{2}\right)\Big(-i \pi  \theta ^2+4 \theta ^2 \Cl(1,\theta )-12 \LS(3,1,\theta )\Big)\\
\tilde F_{8}^{1}(\theta)\quad = \quad &\frac{1}{1440}\csc ^2\left(\frac{\theta }{2}\right)\Big(553 \pi ^4+120 \pi ^2 \theta ^2+840 \pi ^2 \log ^2(2)-840 \log ^4(2)+1440 \LS(2,0,\theta )^2    \nonumber\hfill \\ & 
+\big(1440 \big(\frac{i \pi  \theta }{2}-i \pi ^2\big)+8640 (2 \pi -\theta ) \Cl(1,\theta )\big) \LS(2,0,\theta -\pi )    \nonumber\hfill \\ & 
+\LS(2,0,\theta ) (10080 \LS(2,0,\theta -\pi )-2880 \theta  \Cl(1,\theta ))-5040 \pi  \LS(3,0,\theta -\pi )+4320 \Cl(1,\theta ) \LS(3,1,\theta )    \nonumber\hfill \\ & 
+(17280 \Cl(1,\theta )-1440 i \pi ) \LS(3,1,\theta -\pi )-2880 \LS(4,1,\theta )+1260 \LS(4,1,2 \theta )-5040 \LS(4,1,\theta -\pi )    \nonumber\hfill \\ & 
+10080 \LSLSC(1,1,2,\theta )-20160 \text{Li}_4\left(\frac{1}{2}\right)-360 i \Cl(1,\theta ) \left(\pi  \theta ^2+84 i \zeta (3)\right)    \nonumber\hfill \\ & 
-2520 i \pi  \zeta (3)-17640 \log (2) \zeta (3)\Big)
\end{align}

\subsection{Six propagator integrals}
\begin{align}
\tilde F_{9}^{0}(\theta)\quad = \quad &-\frac{1}{16} \csc ^3\left(\frac{\theta }{2}\right) \sec \left(\frac{\theta }{2}\right) \Big(2 \LS(2,0,\theta ) \theta ^2-22 \LS(3,1,\theta ) \theta -40 \LS(3,1,\theta -\pi ) \theta -70 \zeta (3) \theta+27 \LS(4,2,\theta ) \hfill \nonumber \\ &
+4 \left(2 \theta ^2-10 \pi  \theta +9 \pi ^2\right) \LS(2,0,\theta -\pi )+72 \pi  \LS(3,1,\theta -\pi )+36 \LS(4,2,\theta -\pi )+72 \pi  \zeta (3)\Big)
\end{align}

\begin{align}
\tilde F_{10}^{-1}(\theta)\quad = \quad &\frac{1}{4} \csc ^4\left(\frac{\theta }{2}\right)\Big((2 \theta -4 \pi ) \LS(2,0,\theta -\pi )-4 \LS(3,1,\theta -\pi )-7 \zeta (3)\Big)\\
\tilde F_{10}^{0}(\theta)\quad = \quad &\frac{1}{1440} \csc ^4\left(\frac{\theta }{2}\right)\Big(-158 \pi ^4-240 \pi ^3 \theta -75 \theta ^4-240 \pi ^2 \log ^2(2)+240 \log ^4(2)    \nonumber\hfill \\ & 
+(-1440 \theta -1440 (4 \pi -2 \theta ) \Cl(1,\theta )-1440 \LS(2,0,\theta )+2880 \pi ) \LS(2,0,\theta -\pi )    \nonumber\hfill \\ & 
+2880 \LS(2,0,\theta -\pi )^2+1440 (2 \theta +\pi ) \LS(3,0,\theta -\pi )+(2880-5760 \Cl(1,\theta )) \LS(3,1,\theta -\pi )    \nonumber\hfill \\ & 
+1440 \LS(4,1,\theta )-360 \LS(4,1,2 \theta )+1440 \LS(4,1,\theta -\pi )+1440 \theta  \LSC(2,2,\theta )    \nonumber\hfill \\ & 
-2880 \LSLSC(1,1,2,\theta )+5760 \text{Li}_4\left(\frac{1}{2}\right)+5040 \zeta (3)-10080 \Cl(1,\theta ) \zeta (3)+5040 \log (2) \zeta (3)  
\Big)
\end{align}


\section{Virtual corrections to $gg\rightarrow h$}
\label{sec:hggvertex}
In this section we will present the results for the two loop
corrections to the Higgs boson production process via gluon fusion,
with either quarks or scalar quarks running in the loops, in the region below threshold.
\FIGURE[h]{
\begin{tabular}{ccc}
\fdiagone{37}&\fdiagtwo{37}&\\
\sdiagfive{27}&\sdiagthree{37}&\sdiagfour{37}\\
&&\\
&&
\end{tabular}
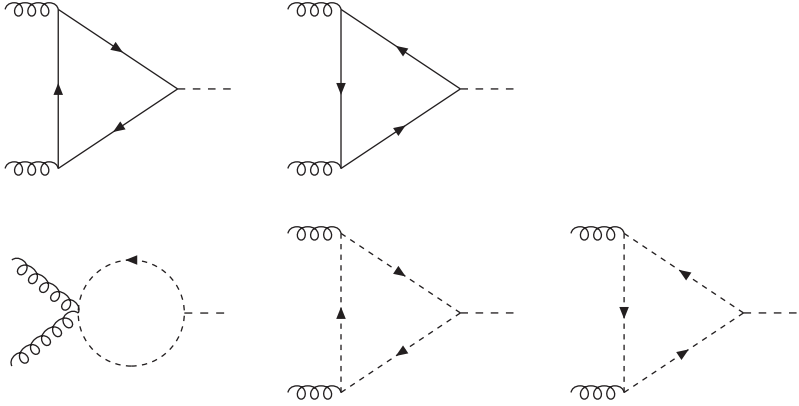
\caption{Contributions to $gg\rightarrow h$ at the lowest order}
\label{fig:born}
}

At order $\alpha_s^2$, the unrenormalized amplitude for the process
$gg\rightarrow H$, is given by
\begin{equation}\label{eq:ampdef}
{\cal M}=\frac{\alpha_s}{4\pi}\,
\sum_{i={s,f}}\left({\cal M}^{(0)}_{i}+\frac{\alpha_s}{4\pi}{\cal M}^{(1)}_{i}\right)
+{\cal O}(\alpha_s^3)\,,
\end{equation}
where $s$ and $f$ denote the contributions of scalars and 
fermions in the loop. The Born amplitudes are given by the diagrams in 
 Figure \ref{fig:born}. They are given by 
\begin{equation}
{\cal M}^{(0)}_{i}=\delta_{ab}\,
K_{ab}\,\Lambda_{i}\,
\left(\frac{m_{i}^2}{4\pi\mu^2}\right)^{-\epsilon}\,e^{-\epsilon\,\gamma_E}\,
M^{(0)}_{i}\,.
\end{equation}
Explicit expressions for the form factors $M^{(0)}_i$  
are given in sections \ref{sec:ggHtop} and \ref{sec:ggHstop} for
fermions and scalars respectively. 
Indices $a$ and $b$ denote the colors of the gluons, 
$\epsilon_i(p_i)$ the corresponding
polarization vectors and $m_i$ the mass of the particle running in the loop.
The couplings of fermions and scalars to the Higgs boson have been written as 
$g_{ffH}=\Lambda_{f}\,m_f$ and $g_{ssH}=\Lambda_{s}\,m_s^2$ respectively.
The constants $\Lambda_{i}$ have inverse mass dimensions, in the case of the SM,
$\Lambda_f=1/v$, where $v$ is the VEV of the Higgs Boson. 
Finally, the helicity projector $K_{ab}$ is given by
\begin{equation}
K_{ab}=\epsilon_a(p_1)\cdot\epsilon_b(p_2)\,p_1\cdot p_2-p_1\cdot\epsilon_b(p_2)\,p_2\cdot
\epsilon_a(p_1)\,.
\end{equation}
The two loop contributions to the amplitudes can be written as 
\begin{eqnarray}
\left(\frac{\alpha_s}{4\pi}\right)^2{\cal M}^{(1)}_{i}=
&&{\cal M}^{(1)}_{i,\text{ir}}+{\cal M}^{(1)}_{i,\text{uv}}
+\delta_{ab}\,K_{ab}\,
\Lambda_{i}\,\left(\frac{\alpha_s}{4\pi}\right)^2\,
\left(M^{(1)}_{i,\text{fin}}+2\,i\,\pi\beta_0\right)+{\cal O}(\epsilon)\,,
\end{eqnarray}
where the infrared and ultraviolet poles have been extracted into
${\cal M}^{(1)}_{i,\text{ir}}$ and ${\cal M}^{(1)}_{i,\text{uv}}$ respectively.
The form factors $M^{(1)}_{i,\text{fin}}$ are finite in the limit 
$\epsilon\rightarrow 0$, their explicit expressions in the region below threshold 
are given in sections \ref{sec:ggHtop} and \ref{sec:ggHstop}.
We have omitted contributions from diagrams with gluon
self-energies in external lines. These drop out if one renormalizes in a heavy quark and squark
decoupling scheme; the running of  the
renormalized strong coupling is then determined from the light flavors
below the decoupling scale \cite{Nason:1989zy,Rodrigo:1993hc}.

The singular contributions can be written in terms of the Born amplitudes. For the
infrared contributions we have
\begin{equation}
{\cal  M}^{(1)}_{i,\text{ir}}=
-\frac{\alpha_s}{4\pi}\left(\frac{-s}{4\pi\mu^2}\right)^{-\epsilon}\,
e^{-\epsilon\,\gamma_E}\,\left[N\,\left(\frac{2}{\epsilon^2}-\frac{\pi^2}{6}\right)
+\frac{2\,\beta_0}{\epsilon}\right]\,{\cal M}^{(0)}_{i}+{\cal O}(\epsilon)\,.
\end{equation}
The ultraviolet pieces, in turn, are given by
\begin{equation}
{\cal M}^{(1)}_{i,\text{uv}}=-\left(\frac{s}{\mu^2}\right)^{-\epsilon}\left(
2\,\delta Z_g\,{\cal M}^{(0)}_{i}+\delta Z_{m_i,\text{gluon}}\,
\frac{\partial}{\partial m_i}\left({\cal M}^{(0)}_{i}\right)\right)\,,
\end{equation}
where $\delta Z_g$ is the strong coupling renormalization constant at one loop in the 
$\overline{{\rm MS}}$ scheme, given by
\begin{equation}
\delta Z_g=-\frac{\alpha_s}{4\pi}\,(4\pi)^{\epsilon}\,e^{-\epsilon\,\gamma_E}\,\frac{\beta_0}{\epsilon}\,,
\end{equation}
and $\delta Z_{m_i,\text{gluon}}$ are the gluonic contributions to the mass renormalization
to fermions and scalars, also in the $\overline{{\rm MS}}$ scheme:
\begin{equation}
\delta Z_{m_i,\text{gluon}}=-\frac{\alpha_s}{4\pi}\,(4\pi)^{\epsilon}
\,e^{-\epsilon\,\gamma_E}\,C_F\,\frac{3}{\epsilon}\,.
\end{equation}
As discussed in the introduction, the fermionic contributions have been computed
in \cite{Spira:1995rr} and expressed in terms of 8 one-dimensional integrals. In turn,
these integrals were computed in terms of harmonic polylogarithms in 
\cite{Harlander:2005rq}, 
giving the analytical result for the two loop corrections with fermions in the loop. 
Our results for these pieces fully agree with the ones quoted in \cite{Harlander:2005rq}.

In the case of scalars mediating the gluon-Higgs boson interaction, 
there are additional contributions originated in quartic couplings 
between the scalars. In supersymmetric theories, these interactions
have a component proportional to $g_s^2$.
Contributions containing the quartic interactions involve,
additionally, the mixing between different scalar quarks. Therefore,
the NLO corrections associated to them contain more than one massive
particle running in the loops.
However, as the gluon couplings to 
the scalars are diagonal, the corrections due to mixing only give contributions
in the form of products of one loop integrals.
In what follows we will only consider the gluonic corrections and postpone the
treatment of contributions due to self interactions of the scalars to
a forthcoming publication.

If the mass of the Higgs particle is significantly smaller than the
mass of the particles circulating in the loops, the amplitudes can 
be approximated by their limit when $m_i\rightarrow \infty$. These results
have been obtained in the context of effective field theories, both for 
fermions and scalar loops in references \cite{Dawson:1990zj,Djouadi:1991tk} 
and \cite{Dawson:1996xz} respectively. Choosing the $\overline{{\rm MS}}$ scheme -notice 
that in the infinite mass limit, the scheme dependence due to mass 
renormalization cancels out- we obtain 
\begin{equation}
{\cal M}_{\text{f},\infty}=\delta_{ab}\,K_{ab}\,\Lambda_f\,
\frac{\alpha_s}{3\,\pi}\,\left(1+\frac{\alpha_s}{\pi}\,
\frac{11}{4}\right)+{\cal M}^{(1)}_{\text{f},\text{ir},\infty}+{\cal O}(\epsilon)\,,
\end{equation}
for fermion loops, whereas for scalar quarks, the amplitude is given by
\begin{equation}
{\cal M}_{\text{s},\infty}=\delta_{ab}\,K_{ab}\,\Lambda_s\,
\frac{\alpha_s}{24\,\pi}\,
\left(1+\frac{\alpha_s}{\pi}\,
\frac{9}{2}\right)+{\cal M}^{(1)}_{\text{s},\text{ir},\infty}+{\cal O}(\epsilon)\,.
\end{equation}
As mentioned above, this last result, which agrees with \cite{hp2spira}, does not include the self
interactions of the scalars, thus it differs from the one quoted in
\cite{Dawson:1996xz}. 
Including a four scalar vertex with coupling 
given by $\sum_a g_s^2\,\left(T^a_{ij}\,T^a_{kl}+T^a_{il}T^a_{jk}\right)$, and
modifying accordingly the mass renormalization of the scalars, we find 
\begin{equation}
{\cal M}_{\text{s},\infty,4}=\delta_{ab}\,K_{ab}\,\Lambda_s\,
\frac{\alpha_s}{24\,\pi}\,
\left(1+\frac{\alpha_s}{\pi}\,
\frac{25}{6}\right)+{\cal M}^{(1)}_{\text{s},\text{ir},\infty}+{\cal O}(\epsilon)\,,
\end{equation}
in agreement with \cite{Dawson:1996xz}. 

Notice that in the expressions quoted above, we have taken the limit $\epsilon\rightarrow 0$,
except in the infrared singular pieces, before evaluating the
limit $m_i\rightarrow \infty$. The infrared contributions contain a prefactor
$(m_i^2)^{-\epsilon}$ that can be expanded only after combining with the
real radiation pieces.

The following two subsections contain the explicit results for the form factors
$M^{(0)}_{i}$ and $M^{(1)}_{i,\text{fin}}$ in the region below threshold, $s<4\,
m_i^2$. As discussed
above, in this region the natural variable to use is $\theta$ defined by
$x=\exp(i\theta)$, with $0\le\theta<\pi$. 
In Appendix \ref{ap:amplitudes} we also give the results
for the form factors as linear combinations of the master integrals introduced in
the previous sections.

\subsection{Amplitudes for quarks}
\label{sec:ggHtop}
At the one loop level, the contributions from fermion
loops to the process $gg\rightarrow H$ when
$s<4\,m_f^2$ are given by
\begin{eqnarray}
M^{(0)}_{\text{f}}=
&&\frac{1}{4}\,(4 - \theta^2 - (4 + \theta^2)\,\cos(\theta))\,
  \csc^{4}\left(\frac{\theta}{2}\right)
\nonumber\\
&&+\epsilon\Bigg[
-2\,(2\,\pi - \theta)\,\cot^{2}\left(\frac{\theta}{2}\right)\,
   \csc^{2}\left(\frac{\theta}{2}\right)\,\LS(2, 0, \theta-\pi ) 
- 4\,\cot^{2}\left(\frac{\theta}{2}\right)\,\csc^{2}\left(\frac{\theta}{2}\right)
  \,\LS(3, 1,\theta -\pi) 
\nonumber\\&&\qquad
+ \frac{1}{2}\csc^{4}\left(\frac{\theta}{2}\right)\,\left(6 - \theta^2 - 2\,\theta\,\sin(\theta) 
- 7\,\zeta(3)(1 + \cos(\theta))-6\,\cos(\theta)\right)\Bigg]
\nonumber\\
&&+\epsilon^2\Bigg[
-4\,\theta\,\cot\left(\frac{\theta}{2}\right)\,
  \csc^{2}\left(\frac{\theta}{2}\right)\,\Cl(1, \theta-\pi) 
+ 2\,\cot^{2}\left(\frac{\theta}{2}\right)\,\csc^{2}\left(\frac{\theta}{2}\right)
  \,\LS(2, 0,  \theta-\pi)^2 
\nonumber\\&&\qquad\,\,
+ 2\,\theta\,\cot^{2}\left(\frac{\theta}{2}\right)\,\csc^{2}\left(\frac{\theta}{2}\right)\,
  \LS(3, 0, \theta-\pi) 
- 4\,\csc^{4}\left(\frac{\theta}{2}\right)\,\LS(3, 1, \theta-\pi) 
\nonumber\\&&\qquad\,\,
- 2\,\csc^{4}\left(\frac{\theta}{2}\right)\,(2\,\pi - \theta - \sin(\theta))
  \,\LS(2, 0, \theta-\pi)
+\frac{1}{48}\csc^{4}\left(\frac{\theta}{2}\right)\,\big(336 + 4\,\pi^2 - 4\,\pi^3\,\theta 
- 24\,\theta^2
\nonumber\\&&\qquad\,\,
   - \pi^2\,\theta^2 
- (336 + 4\,\pi^3\,\theta + \pi^2\,(4 + \theta^2))
  \,\cos(\theta) - 144\,\theta\,\sin(\theta) - 336\,\zeta(3)\big)\Bigg]+{\cal O}(e^3)\,.

\end{eqnarray}
The form factor at two loops, in turn, is given by:
\begin{eqnarray}
M^{(1)}_{\text{f},\text{fin}}=
\frac{1}{N}\,\Bigg[&&
\frac{27}{2}\,\cos\left(\frac{\theta}{2}\right)\,\cos(\theta)
  \,\LS(4, 2, \theta)
+ 18\,\cos\left(\frac{\theta}{2}\right)\,\cos(\theta)
  \,\LS(4, 2, \theta-\pi)
\nonumber\\&&
+ \theta\,\cos\left(\frac{\theta}{2}\right)
  \,(\theta\,\cos(\theta)- \sin(\theta))\,\LS(2, 0, \theta) 
\nonumber\\&&
 + 4\,\cos\left(\frac{\theta}{2}\right)
  \,((9\,\pi - 5\,\theta)\,\cos(\theta) + 2\,\sin(\theta))
  \,\LS(3, 1, \theta-\pi)
\nonumber\\&&
+ 2\,\cos\left(\frac{\theta}{2}\right)
  \,((9\,\pi^2 - 10\,\pi\,\theta + 2\,\theta^2)
  \,\cos(\theta) + 2\,(2\,\pi - \theta)\,\sin(\theta))\,\LS(2, 0,\theta -\pi)
\nonumber\\&&
- \,\sin\left(\frac{\theta}{2}\right)\,
  ((6 + 4\,\theta^2)\,\cos(\theta) + 3\,(-2 + \theta^2 - \theta\,\sin(\theta)))\,\Cl(1, \theta)
\nonumber\\&&
- \frac{1}{2}\,\Bigg(11\,\theta\,\cos\left(\frac{\theta}{2}\right) 
  + 11\,\theta\,\cos\left(\frac{3\theta}{2}\right) 
  + 9\,\sin\left(\frac{\theta}{2}\right) 
  - 7\,\sin\left(\frac{3\theta}{2}\right)\Bigg)\,\LS(3, 1, \theta)
\nonumber\\&&
+ \frac{1}{8}\,\Bigg(\cos\left(\frac{\theta}{2}\right)
  \,\left(-3\,\theta^3 + \theta\,(4 - 140\,\zeta(3)) + 144\,\pi\,\zeta(3)\right) 
\nonumber\\&&\qquad
- 4\,\sin\left(\frac{\theta}{2}\right)\,(18 - 9\,\theta^2 - 28\,\zeta(3) - 
2\,\cos(\theta)\,(9 + \theta^2 + 14\,\zeta(3))) 
\nonumber\\&&\qquad
- \cos\left(\frac{3\theta}{2}\right)
  \,\left(\theta^3 - 144\,\pi\,\zeta(3) + 4\,\theta\,(1 + 35\,\zeta(3))\right)\Bigg)\Bigg]
\,\csc^{5}\left(\frac{\theta}{2}\right)
\nonumber\\
-N\,\Bigg[&&
- 16\,\cos^{2}\left(\frac{\theta}{2}\right)\,\sin\left(\frac{\theta}{2}\right) 
  \,\LSLSC(1, 1, 2, \theta)
\nonumber\\&&
+\frac{27}{2}\,\cos\left(\frac{\theta}{2}\right)\,\cos(\theta)
  \,\LS(4, 2, \theta)
+ 18\,\cos\left(\frac{\theta}{2}\right)\,\cos(\theta) 
  \,\LS(4, 2, \theta-\pi)
\nonumber\\&&
+ 8\,\pi\,\cos^{2}\left(\frac{\theta}{2}\right)\,\sin\left(\frac{\theta}{2}\right)
  \,\LS(3, 0,\theta -\pi)
+ 8\,\cos^{2}\left(\frac{\theta}{2}\right)\,\sin\left(\frac{\theta}{2}\right) 
\,\LS(4, 1, \theta)
\nonumber\\&&
- 2\,\cos^{2}\left(\frac{\theta}{2}\right)\,\sin\left(\frac{\theta}{2}\right) 
  \,\LS(4, 1, 2\,\theta)
+ 8\,\cos^{2}\left(\frac{\theta}{2}\right)\,\sin\left(\frac{\theta}{2}\right) 
  \,\LS(4, 1, \theta-\pi)
\nonumber\\&&
- \bigg(16\,\cos^{2}\left(\frac{\theta}{2}\right)\,
\sin\left(\frac{\theta}{2}\right)\,\LS(2, 0, \theta-\pi)
   - \theta\,\cos\left(\frac{\theta}{2}\right)
  \,(\theta\,\cos(\theta) - \sin(\theta))\bigg)
  \,\LS(2, 0, \theta)
\nonumber\\&&
- 4\,\bigg(8\,\Cl(1, \theta)\,\cos^{2}\left(\frac{\theta}{2}\right)
  \,\sin\left(\frac{\theta}{2}\right) 
\nonumber\\&&
- \cos\left(\frac{\theta}{2}\right)
  \,((9\,\pi - 5\,\theta)\,\cos(\theta) + 2\,\sin(\theta))\bigg)\,\LS(3, 1,  \theta-\pi)
\nonumber\\&&
-2\,\bigg(8\,(2\,\pi - \theta)\,\Cl(1, \theta)\,\cos^{2}\left(\frac{\theta}{2}\right)
  \,\sin\left(\frac{\theta}{2}\right) 
\nonumber\\&&
  - \cos\left(\frac{\theta}{2}\right)
  \,\left((9\,\pi^2 - 10\,\pi\,\theta + 2\,\theta^2)\,\cos(\theta) 
  + 2\,(2\,\pi - \theta)\,\sin(\theta)\right)\bigg)\, \LS(2, 0, \theta-\pi)
\nonumber\\&&
- \frac{1}{2}\,
 \bigg(11\,\theta\,\cos\left(\frac{\theta}{2}\right) 
  + 11\,\theta\,\cos\left(\frac{3\theta}{2}\right) 
  - 5\,\sin\left(\frac{\theta}{2}\right) 
  - 13\,\sin\left(\frac{3\theta}{2}\right)\bigg)\,\LS(3, 1, \theta)
\nonumber\\&&
- \sin\left(\frac{\theta}{2}\right)\,\left(-6 + 8\,\theta^2 - 3\,\theta\,\sin(\theta) 
 + 28\,\zeta(3) + \cos(\theta)\,(6 + 7\,\theta^2 + 28\,\zeta(3))\right)\,\Cl(1, \theta)
\nonumber\\&&
- \frac{1}{8}
  \cos\left(\frac{\theta}{2}\right)\,(9\,\theta^3 
     - 4\,\theta\,(1 - 35\,\zeta(3)) - 144\,\pi\,\zeta(3)) 
\nonumber\\&&
 - \frac{1}{8}\,\cos\left(\frac{3\theta}{2}\right)\,(3\,\theta^3  
     + 4\,\theta\,(1 + 35\,\zeta(3))- 144\,\pi\,\zeta(3))
\nonumber\\&&
 - \sin\left(\frac{\theta}{2}\right)\,\Big( 
-\frac{7}{2}\,\theta^2  + 30 
+ \cos^{2}\left(\frac{\theta}{2}\right)\,\big(\frac{79}{90}\,\pi^4 + \frac{4}{3}\,\pi^2\,\log(2)^2 
  -3\,\theta^2 + \frac{1}{4}\,\theta^4 
\nonumber\\&&
  - \frac{2}{3}\,(45 + 2\,\log(2)^4 + 48\,\text{Li}_4(1/2) 
  + 42\,\zeta(3) + 42\,\log(2)\,\zeta(3))\big)\Big)
\Bigg]\,\csc^{5}\left(\frac{\theta}{2}\right)+{\cal O}(\epsilon)\,.

\end{eqnarray}

\subsection{Amplitudes for scalar quarks}
\label{sec:ggHstop}
The one loop form factor for scalar quarks,
$s<4\,m_f^2$ is given by
\begin{eqnarray}
M^{(0)}_{\text{s}}=
&&-\frac{1}{8}(2 - \theta^2 - 2\,\cos(\theta))\,\csc^{4}\left(\frac{\theta}{2}\right)
\nonumber\\
&&+\frac{\epsilon}{8}\Big[4(2\,\pi - \theta)\,\csc^{4}\left(\frac{\theta}{2}\right)
  \,\LS(2, 0,  \theta-\pi)
+ 8\csc^{4}\left(\frac{\theta}{2}\right)\,\LS(3, 1,\theta -\pi ) 
\nonumber\\
&&\qquad
+ \csc^{4}\left(\frac{\theta}{2}\right)\,(-6 + \theta^2 + 6\,\cos(\theta) 
  + 2\,\theta\,\sin(\theta) + 14\,\zeta(3))\Bigg]
\nonumber\\
&&+\frac{\epsilon^2}{96}\Bigg[
96\,\theta\,\cot\left(\frac{\theta}{2}\right)
  \,\csc^{2}\left(\frac{\theta}{2}\right)\,\Cl(1,\theta -\pi)
- 48\,\csc^{4}\left(\frac{\theta}{2}\right)\,\LS(2, 0, \theta-\pi)^2
\nonumber\\
&&\qquad\,\,\,
- 48\,\theta\,\csc^{4}\left(\frac{\theta}{2}\right)\,\LS(3, 0, \theta-\pi)
+ 96\,\csc^{4}\left(\frac{\theta}{2}\right)\,\LS(3, 1, \theta-\pi) 
\nonumber\\
&&\qquad\,\,\,
- 48\,\csc^{4}\left(\frac{\theta}{2}\right)\,(-2\,\pi + \theta + \sin(\theta))
\,\LS(2, 0,  \theta-\pi) 
- \csc^{4}\left(\frac{\theta}{2}\right)\,(168 + 2\,\pi^2 - 4\,\pi^3\,\theta - 12\,\theta^2 
\nonumber\\
&&\qquad\,\,\,
- \pi^2\,\theta^2 - 2\,(84 + \pi^2)\,\cos(\theta) - 72\,\theta\,\sin(\theta) 
- 168\,\zeta(3))\Bigg]+{\cal O}(\epsilon^3)\,.
\end{eqnarray}
The form factor at two loops, in turn is given by:
\begin{eqnarray}
M^{(1)}_{\text{s},\text{fin}}=
-\frac{1}{N}\,\Bigg[
&&
\frac{27}{4}\,\cos(\theta)
\,\LS(4, 2, \theta)
+ 9\,\cos(\theta)
\,\LS(4, 2, \theta-\pi)
\nonumber\\&&
- \frac{1}{4}\,\sin\left(\frac{\theta}{2}\right)
  \,((-3 + 8\,\theta^2)\,\cos\left(\frac{\theta}{2}\right) 
  + 3\,(\cos\left(\frac{3\theta}{2}\right) 
  - 2\,\theta\,\sin\left(\frac{\theta}{2}\right)))\,\Cl(1, \theta)
\nonumber\\&&
- \frac{1}{2}
  \,(11\,\theta\,\cos(\theta) - 3\,\sin(\theta))\,\LS(3, 1, \theta)
+ \frac{1}{2}\,\theta
  \,(\theta\,\cos(\theta) - \sin(\theta))\,\LS(2, 0, \theta)
\nonumber\\&&
+ 2
  \,((9\,\pi - 5\,\theta)\,\cos(\theta) + 2\,\sin(\theta))\,\LS(3, 1,\theta -\pi)
\nonumber\\&&
+ ((9\,\pi^2 - 10\,\pi\,\theta + 2\,\theta^2)\,\cos(\theta) 
 + 2\,(2\,\pi - \theta)\,\sin(\theta))\,\LS(2, 0, \theta-\pi)
\nonumber\\&&
+ \frac{1}{16}\,
  (-11\,\theta + \theta^3 - 3\,\theta\,\cos(2\,\theta) - 28\,\sin(\theta) + 17\,\theta^2\,\sin(\theta) + 14\,\sin(2\,\theta) 
\nonumber\\&&\qquad
+ 112\,\sin(\theta)\,\zeta(3) + \cos(\theta)\,(\theta^3 + \theta\,(14 - 280\,\zeta(3)) + 288\,\pi\,\zeta(3)))
\Bigg]\,\frac{\csc^{5}\left(\frac{\theta}{2}\right)}{2\,\cos\left(\frac{\theta}{2}\right)}
\nonumber\\
-N\,\Bigg[&&
8\,\cos\left(\frac{\theta}{2}\right)\,\sin\left(\frac{\theta}{2}\right)
  \,\LSLSC(1, 1, 2, \theta)
\nonumber\\&&
-\frac{27}{4}\,\cos(\theta)\,\LS(4, 2, \theta)
- 9\,\cos(\theta)\,\LS(4, 2, \theta -\pi) 
\nonumber\\&&
- 4\,\pi\,\cos\left(\frac{\theta}{2}\right)\,\sin\left(\frac{\theta}{2}\right)
  \,\LS(3, 0, \theta-\pi) 
- 4\,\cos\left(\frac{\theta}{2}\right)\,\sin\left(\frac{\theta}{2}\right)
  \,\LS(4, 1, \theta) 
\nonumber\\&&
+ \cos\left(\frac{\theta}{2}\right)\,\sin\left(\frac{\theta}{2}\right) 
  \,\LS(4, 1, 2\,\theta)
- 4\,\cos\left(\frac{\theta}{2}\right)\,\sin\left(\frac{\theta}{2}\right)
  \,\LS(4, 1,\theta -\pi)
\nonumber\\&&
- \frac{1}{2}\,\theta\,(\theta\,\cos(\theta) - \sin(\theta))\,\LS(2, 0, \theta)
+ \frac{11}{2}\,(\theta\,\cos(\theta) - \sin(\theta))\,\LS(3, 1, \theta) 
\nonumber\\&&
+ 2\,\bigg(4\,\Cl(1, \theta)\,\sin(\theta) 
- ((9\,\pi - 5\,\theta)\,\cos(\theta) + 2\,\sin(\theta))\bigg)\,\LS(3, 1, \theta-\pi)
\nonumber\\&&
+\bigg(4\,(2\,\pi - \theta)\,\Cl(1, \theta)\,
  \sin(\theta)
+ 4\,\sin(\theta)\,\LS(2, 0, \theta)
\nonumber\\&&
- ((9\,\pi^2 - 10\,\pi\,\theta + 2\,\theta^2)\,\cos(\theta) 
  + 2\,(2\,\pi - \theta)\,\sin(\theta))\bigg)
  \,\LS(2, 0, \theta-\pi)
\nonumber\\&&
 + \frac{1}{4}\,\sin\left(\frac{\theta}{2}\right)\,\bigg(3\,\cos\left(\frac{3\theta}{2}\right) 
  - 6\,\theta\,\sin\left(\frac{\theta}{2}\right) 
- \cos\left(\frac{\theta}{2}\right)
  \,(3 - 24\,\theta^2 - 112\,\zeta(3))\bigg)\,\Cl(1, \theta)
\nonumber\\&&
 + \frac{1}{16}
\,\bigg(11\,\theta + 3\,\theta^3 + 3\,\theta\,\cos(2\,\theta) + 52\,\sin(\theta) 
+ \frac{158}{45}\,\pi^4\,\sin(\theta) 
\nonumber\\&&
- 21\,\theta^2\,\sin(\theta) + \theta^4\,\sin(\theta) 
 + \frac{16}{3}\,\pi^2\,\log^{2}(2)\,\sin(\theta) - \frac{16}{3}\,\log^{4}(2)\,\sin(\theta) 
\nonumber\\&&
- 128\,\text{Li}_4\left(\frac{1}{2}\right)\,\sin(\theta) - 26\,\sin(2\,\theta) 
- 112\,\sin(\theta)\,\zeta(3) - 112\,\log(2)\,\sin(\theta)\,\zeta(3) 
\nonumber\\&&
+ \cos(\theta)\,(3\,\theta^3 - 14\,\theta\,(1 - 20\,\zeta(3)) 
- 288\,\pi\,\zeta(3))\bigg)\Bigg]
\,\frac{\csc^{5}\left(\frac{\theta}{2}\right)}{2\,\cos\left(\frac{\theta}{2}\right)}+{\cal O}(\epsilon)\,.

\end{eqnarray}

\section{Summary}
\label{sec:conclusions}

In this paper  we have computed the two-loop 
master integrals which are needed for the evaluation of the 
two-loop QCD amplitude in the gluon fusion process $gg \to h$.  
This  is  a loop induced process and generally requires a new 
calculation when modifying the particle content in the loops. 

We have automatized the evaluation of the two-loop amplitude using 
modern reduction methods and providing analytic expressions 
for the master integrals. We computed the master integrals using the 
differential equation method. Our results agree with the literature 
when a comparison is available and  with a direct numerical evaluation
which is performed with an independent method.  

In this paper we have evaluated analytically the two-loop amplitudes 
for $gg \to h$ via a quark and a scalar quark. The first result agrees with the
result of Spira et al., in the analytic form written by Harlander 
and Kant.  The amplitude for the scalar quark is a new result, and agrees with
the result derived within the heavy squark approximation. 

The master integrals we have presented here, are relevant for other 
$2 \to 1$ processes in the Standard Model and its extensions and more 
complicated $2 \to 2$ processes with heavy particles in the loops.

\acknowledgments 
We gratefully thank Uli Haisch for very enlightening and useful discussions and suggestions. 
C.A. is grateful to ETH Z\"urich for their hospitality and providing 
access to MAPLE v10. The work of A.D. was partially supported by the Swiss National Science Foundation
(SNF) under contract number 200020-109162 and by the Forschungskredit der Universit\"at Z\"urich.

\appendix
\section{Amplitudes in terms of master integrals}
\label{ap:amplitudes}
We present the results for the amplitudes in eq. (\ref{eq:ampdef}) in terms
of the master integrals in Section \ref{sec:master}. We write the amplitudes as
\begin{equation}
{\cal M}^{(n)}_{i}=\delta_{ab}\,K_{ab}\,\Lambda_i\,(4\,\pi\,\mu^2)^{(n+1)\,\epsilon}\,
{\cal \bar{M}}^{(n)}_i\,.
\end{equation}
\subsection{Fermionic amplitudes}
For the fermionic contribution at one loop, we have
\begin{equation}
{\cal \bar{M}}^{(0)}_{\text{f}}=
\Bigg\{
\frac{8\,\epsilon\,(1 + \epsilon)\,x}{(1 - x)^2}\bubble{12}-\frac{4\,s\,x\,(1 + (2 + 4\,\epsilon + 4\,\epsilon^2)\,x + x^2)}{(1 - x)^4}\trianb{23}\Bigg\}\,.
\end{equation}
At two loops:
\begin{align}
{\cal \bar{M}}^{(1)}_{\text{f}}=
&\Bigg\{\frac{N}{\epsilon\,s^2\,x^2}\bigg[
-24\,x\,(1 + x)^2 - \epsilon\,(1 + 56\,x + 238\,x^2 + 56\,x^3 + x^4) 
\nonumber\\&\qquad
+ \epsilon^2\,(9 - 360\,x - 994\,x^2 - 360\,x^3 + 9\,x^4) 
+ 2\,\epsilon^3\,(19 - 900\,x - 2014\,x^2 - 900\,x^3 + 19\,x^4)\bigg]
\nonumber\\&\quad+\frac{C_F}{s^2\,x^2\,(1 + x)^2}\bigg[
4\,(1 - 12\,x - 25\,x^2 + 8\,x^3 - 25\,x^4 - 12\,x^5 + x^6) 
\nonumber\\&\qquad
- 8\,\epsilon\,(1 + 10\,x + 51\,x^2 + 36\,x^3 + 51\,x^4 + 10\,x^5 + x^6) 
\nonumber\\&\qquad
- 8\,\epsilon^2\,(5 + 37\,x + 203\,x^2 + 214\,x^3 + 203\,x^4 + 37\,x^5 + 5\,x^6)
\bigg]\Bigg\}{\SetScale{0.8} \dt{22}}
\nonumber\\&+\Bigg\{\frac{N}{s\,(1 - x)^2}\bigg[
24\,(1 + x)^2 + 20\,\epsilon^2\,(17 + 46\,x + 17\,x^2) 
+ 4\,\epsilon\,(23 + 42\,x + 23\,x^2)\bigg]
\nonumber\\&\quad+\frac{C_F}{s\,(1 - x)^2\,(1 + x)^2}\bigg[
8\,(1 - x)^2\,(1 - 6\,x + x^2) 
+ 8\,\epsilon\,(9 - 8\,x + 30\,x^2 - 8\,x^3 + 9\,x^4) 
\nonumber\\&\qquad
+ 8\,\epsilon^2\,(21 + 8\,x + 102\,x^2 + 8\,x^3 + 21\,x^4)\bigg]\Bigg\}\bttwo{17}
\nonumber\\&+\frac{N}{(1 - x)^2}\bigg[
16\,x - 16\,\epsilon\,x - 16\,\epsilon^2\,x\bigg]\db{27}
\nonumber\\&+\Bigg\{\frac{N}{\epsilon^2\,(1 - x)^4}\bigg[
4\,s\,x\,(1 + x)^2 + 2\,\epsilon\,s\,x\,(3 + x)\,(1 + 3\,x) + \epsilon^2\,s\,(1 + 2\,x + 122\,x^2 + 2\,x^3 + x^4)\bigg]
\nonumber\\&\quad+\frac{C_F}{\epsilon\,(1 - x)^4}\bigg[
8\,s\,x\,(1 + x)^2 - 4\,\epsilon\,s\,(1 - 8\,x - 10\,x^2 - 8\,x^3 + x^4)\bigg]\Bigg\}\ssonetwotwo{17}
\nonumber\\&+\Bigg\{\frac{N}{(1 - x)^4}\bigg[
4\,\epsilon\,s\,(1 + x)^2\,(1 - 26\,x + x^2)\bigg]
+\frac{C_F}{(1 - x)^2}\bigg[
-16\,\epsilon\,s\,(1 + x)^2\bigg]\Bigg\}\sstwoonetwo{17}
\nonumber\\&+\Bigg\{\frac{N}{\epsilon\,(1 - x)^4}\bigg[
-24\,(1 + x)^2\,(1 - 4\,x + x^2) 
- 4\,\epsilon\,(1 - 4\,x + x^2)\,(7 + 26\,x + 7\,x^2)
\nonumber\\&\qquad
- 4\,\epsilon^2\,(43 - 46\,x - 442\,x^2 - 46\,x^3 + 43\,x^4) 
- 4\,\epsilon^3\,(1 - 4\,x + x^2)\,(201 + 550\,x + 201\,x^2) \bigg]
\nonumber\\&\quad+\frac{C_F}{(1 - x)^4}\bigg[
-16\,(1 + x)^2\,(1 - 4\,x + x^2) 
- 8\,\epsilon\,(7 - 16\,x - 30\,x^2 - 16\,x^3 + 7\,x^4) 
\nonumber\\&\qquad
- 16\,\epsilon^2\,(10 - 19\,x - 70\,x^2 - 19\,x^3 + 10\,x^4)\bigg]\Bigg\}\tria{23}
\nonumber\\&+\Bigg\{\frac{N}{\epsilon^2\,(1 - x)^6}\bigg[
-32\,s^2\,x^2\,(1 + x)^2 - 16\,\epsilon\,s^2\,x^2\,(5 + 14\,x + 5\,x^2) - 8\,\epsilon^2\,s^2\,x^2\,(41 + 118\,x + 41\,x^2)\bigg]
\nonumber\\&\quad+\frac{C_F}{\epsilon\,(1 - x)^6}\bigg[
-16\,s^2\,x^2\,(1 + x)^2 - 32\,\epsilon\,s^2\,x^2\,(3 + 4\,x + 3\,x^2)\bigg]\Bigg\}\triathree{23}
\nonumber\\&+\Bigg\{\frac{N}{\epsilon\,(1 - x)^4}\bigg[
-8\,s\,x\,(1 + x)^2 + 16\,\epsilon\,s\,x\,(1 + x^2)\bigg]\Bigg\}\dtria{23}
\nonumber\\&+\Bigg\{\frac{N}{(1 - x)^2}\bigg[
32\,\epsilon\,(1 + x)^2 + 4\,(1 + 6\,x + x^2)\bigg]
\nonumber\\&\quad
-\frac{C_F}{(1 - x)^2}\bigg[
 8\,(1 + 14\,x + x^2)+96\,\epsilon\,x \bigg]\Bigg\}\tritad{23}
\nonumber\\&+\Bigg\{\frac{N}{(1 - x)^2}\bigg[
-80\,\epsilon^2\,(1 + x)^2 + 2\,(1 - 18\,x + x^2) - 16\,\epsilon\,(1 + 4\,x + x^2)\bigg]
\nonumber\\&\quad+\frac{C_F}{(1 - x)^2}\bigg[
-8\,(1 - 6\,x + x^2) + 8\,\epsilon\,(1 + 6\,x + x^2) + 8\,\epsilon^2\,(5 + 6\,x + 5\,x^2)\bigg]\Bigg\}\mpfour{23}
\nonumber\\&+\quad\frac{C_F}{(1 - x)^2\,(1 + x)^2}\bigg[
-32\,\epsilon^2\,x^2 + 16\,x\,(1 + x^2) - 16\,\epsilon\,x\,(1 + 4\,x + x^2)\bigg]\glasses{17}
\nonumber\\&+\Bigg\{\frac{N}{(1 - x)^4}\bigg[
8\,s\,x\,(1 + x)^2 + 24\,\epsilon\,s\,x\,(1 + x)^2\bigg]
\nonumber\\&\quad+\frac{C_F}{(1 - x)^4}\bigg[
-16\,\epsilon\,s\,(1 - x)^2\,x + 8\,s\,x\,(1 + x)^2\bigg]\Bigg\}\tribub{23}
\nonumber\\&+\frac{C_F}{(1 - x)^6}\bigg[
-8\,s^2\,x\,(1 + x)^2\,(1 + x^2)\bigg]\mpsix{23}
\nonumber\\&+\Bigg\{\frac{N}{(1 - x)^4}\bigg[
-4\,s^2\,x\,(1 + x)^2 - 2\,\epsilon\,s^2\,x\,(3 + x)\,(1 + 3\,x)\bigg]\Bigg\}\xtria{23}
\nonumber\\&+\Bigg\{\frac{N}{\epsilon\,s\,(1 - x)^2}\bigg[
-96\,(1 + x)^2 - 32\,\epsilon\,(5 + 16\,x + 5\,x^2) 
\nonumber\\&\quad
- 8\,\epsilon^2\,(105 + 326\,x + 105\,x^2) - 24\,\epsilon^3\,(161 + 446\,x + 161\,x^2)\bigg]
\nonumber\\&\quad-\frac{16\,C_F}{s\,(1 - x)^2}\bigg[
4\,(1 + x)^2 + 8\,\epsilon\,(2 + 3\,x + 2\,x^2) 
\nonumber\\&\quad
+ 3\,\epsilon^2\,(17 + 38\,x + 17\,x^2)\bigg]\Bigg\}\triatwo{23}+
{\cal O}(\epsilon)\,.
\end{align}

\subsection{Scalar amplitudes}
The scalar contributions are given by
\begin{equation}
{\cal \bar{M}}^{(0)}_{\text{s}}=-\frac{2\,\epsilon\,(1 + \epsilon)\,x}{(1 - x)^2}\bubble{12}
+\frac{4\,(1 + \epsilon + \epsilon^2)\,s\,x^2}{(1 - x)^4}\trianb{23}\,,
\end{equation}
at one loop, and
\begin{align}
{\cal \bar{M}}^{(1)}_{\text{s}}=
&+\Bigg\{\frac{N}{\epsilon\,s^2\,x}\bigg[
24\,x - 4\,\epsilon^2\,(1 - 108\,x + x^2) - 4\,\epsilon\,(1 - 24\,x + x^2) + 4\,\epsilon^3\,(5 + 462\,x + 5\,x^2)\bigg]
\nonumber\\&\quad+\frac{C_F}{s^2\,x\,(1 + x)^2}\bigg[
4\,x\,(9 - 2\,x + 9\,x^2) 
+ 4\,\epsilon\,(1 + 23\,x + 32\,x^2 + 23\,x^3 + x^4) 
\nonumber\\&\qquad
+ 8\,\epsilon^2\,(1 + 48\,x + 78\,x^2 + 48\,x^3 + x^4)\bigg]\Bigg\}\dt{27}
\nonumber\\&+\Bigg\{\frac{N}{s\,(1 - x)^2}\bigg[
-24\,x - 88\,\epsilon\,x - 400\,\epsilon^2\,x\bigg]
\nonumber\\&\quad+\frac{C_F}{s\,(1 - x)^2\,(1 + x)^2}\bigg[
 - 4\,(1 - x)^2\,(2 - 3\,x + 2\,x^2) 
- 8\,\epsilon\,(1 + 3\,x + 3\,x^3 + x^4)
\nonumber\\&\qquad
-16\,\epsilon^2\,(1 + x)^2\,(1 + 3\,x + x^2)\bigg]\Bigg\}\bttwo{17}
\nonumber\\&+\frac{N}{(1 - x)^2}\bigg[
-4\,x + 4\,\epsilon\,x + 4\,\epsilon^2\,x\bigg]\db{27}
\nonumber\\&+\Bigg\{\frac{N}{\epsilon^2\,(1 - x)^4}\bigg[
-4\,s\,x^2 - 8\,\epsilon\,s\,x^2 + 4\,\epsilon^2\,s\,x\,(1 - 10\,x + x^2)\bigg]
\nonumber\\&\quad+\frac{C_F}{\epsilon\,(1 - x)^4}\bigg[
-8\,s\,x^2 - 24\,\epsilon\,s\,x^2\bigg]\Bigg\}\ssonetwotwo{17}
\nonumber\\&+\frac{N}{(1 - x)^4}\bigg[
24\,\epsilon\,s\,x\,(1 + x)^2\bigg]\sstwoonetwo{17}
\nonumber\\&+\Bigg\{\frac{N}{\epsilon\,(1 - x)^4}\bigg[
24\,x\,(1 - 4\,x + x^2) + 40\,\epsilon\,x\,(1 - 4\,x + x^2)  
+ 8\,\epsilon^2\,x\,(27 - 110\,x + 27\,x^2)
\nonumber\\&\qquad
+ 952\,\epsilon^3\,x\,(1 - 4\,x + x^2)\bigg]
\nonumber\\&\quad+\frac{C_F}{(1 - x)^4}\bigg[
16\,x\,(1 - 4\,x + x^2) + 8\,\epsilon\,x\,(5 - 22\,x + 5\,x^2) 
\nonumber\\&\quad
+ 8\,\epsilon^2\,x\,(21 - 86\,x + 21\,x^2)\bigg]\Bigg\}\tria{23}
\nonumber\\&+\Bigg\{\frac{N}{\epsilon^2\,(1 - x)^6}\bigg[
32\,s^2\,x^3 + 96\,\epsilon\,s^2\,x^3 + 400\,\epsilon^2\,s^2\,x^3\bigg]
\nonumber\\&\quad
+\frac{C_F}{\epsilon\,(1 - x)^6}\bigg[
16\,s^2\,x^3 + 80\,\epsilon\,s^2\,x^3\bigg]\Bigg\}\triathree{23}
\nonumber\\&+\frac{N}{\epsilon\,(1 - x)^4}\bigg[
8\,s\,x^2 - 8\,\epsilon\,s\,x^2\bigg]\dtria{23}
\nonumber\\&+\Bigg\{\frac{N}{(1 - x)^2}\bigg[
-8\,x - 32\,\epsilon\,x\bigg]
+\frac{C_F}{(1 - x)^2}\bigg[
20\,x + 32\,\epsilon\,x\bigg]\Bigg\}\tritad{23}
\nonumber\\&+\Bigg\{\frac{N}{(1 - x)^2}\bigg[
8\,x + 24\,\epsilon\,x + 80\,\epsilon^2\,x\bigg]
+\frac{C_F}{(1 - x)^2}\bigg[
-8\,x - 16\,\epsilon\,x - 32\,\epsilon^2\,x\bigg]\Bigg\}\mpfour{23}
\nonumber\\&+\frac{C_F}{(1 - x)^2\,(1 + x)^2}\bigg[
2\,\epsilon\,(1 - x)^2\,x + 2\,\epsilon^2\,(1 - x)^2\,x - 4\,x\,(1 + x^2)\bigg]
\glasses{17}
\nonumber\\&+\Bigg\{\frac{N}{(1 - x)^4}\bigg[
-8\,s\,x^2 - 24\,\epsilon\,s\,x^2\bigg]
\nonumber\\&\quad+\frac{C_F}{(1 - x)^4}\bigg[
4\,s\,x^2 + 4\,\epsilon\,s\,x^2\bigg]\Bigg\}\tribub{23}
\nonumber\\&+\frac{C_F}{(1 - x)^6}\bigg[
8\,s^2\,x^2\,(1 + x^2)\bigg]\mpsix{23}
\nonumber\\&+\Bigg\{\frac{N}{(1 - x)^4}\bigg[
s\,x^2\bigg]
+\frac{C_F}{(1 - x)^4}\bigg[
-4\,s\,x^2\bigg]\Bigg\}\mpfive{23}
\nonumber\\&+\frac{N}{(1 - x)^4}\bigg[
4\,s^2\,x^2 + 8\,\epsilon\,s^2\,x^2\bigg]
\xtria{23}
\nonumber\\&+\Bigg\{\frac{N}{\epsilon\,s\,(1 - x)^2}\bigg[
96\,x + 208\,\epsilon\,x + 1072\,\epsilon^2\,x + 4608\,\epsilon^3\,x\bigg]
\nonumber\\&\quad+\frac{C_F}{s\,(1 - x)^2}\bigg[
64\,x + 224\,\epsilon\,x + 864\,\epsilon^2\,x\bigg]\Bigg\}\triatwo{23}
+{\cal O}(\epsilon)\,,
\end{align}
at two loops.
%


\begin{thebibliography}{99}
\bibitem{pittau}
  G.~Ossola, C.~G.~Papadopoulos and R.~Pittau,
  arXiv:hep-ph/0609007.


\bibitem{manohar}
  A.~V.~Manohar and M.~B.~Wise,
  Phys.\ Lett.\ B {\bf 636}, 107 (2006)
  [arXiv:hep-ph/0601212].


\bibitem{Spira:1995rr}
M.~Spira, A.~Djouadi, D.~Graudenz and P.~M.~Zerwas,
Phys.\ Lett.\ B {\bf 318}, 347 (1993);
  M.~Spira, A.~Djouadi, D.~Graudenz and P.~M.~Zerwas,
  Nucl.\ Phys.\ B {\bf 453}, 17 (1995)
  [arXiv:hep-ph/9504378].

\bibitem{Dawson:1990zj}
  S.~Dawson,
  Nucl.\ Phys.\ B {\bf 359}, 283 (1991).

\bibitem{laporta}
  S.~Laporta,
  Int.\ J.\ Mod.\ Phys.\ A {\bf 15}, 5087 (2000)
  [arXiv:hep-ph/0102033].

\bibitem{smirnov}
  V.~A.~Smirnov and O.~L.~Veretin,
  Nucl.\ Phys.\ B {\bf 566}, 469 (2000)
  [arXiv:hep-ph/9907385].


\bibitem{tausk}
  J.~B.~Tausk,
  Phys.\ Lett.\ B {\bf 469}, 225 (1999)
  [arXiv:hep-ph/9909506].

\bibitem{tausk1}
  C.~Anastasiou, J.~B.~Tausk and M.~E.~Tejeda-Yeomans,
  Nucl.\ Phys.\ Proc.\ Suppl.\  {\bf 89}, 262 (2000)
  [arXiv:hep-ph/0005328].

\bibitem{ourmb}
  C.~Anastasiou and A.~Daleo,
  JHEP {\bf 0610}, 031 (2006)
  [arXiv:hep-ph/0511176].

\bibitem{czakonmb}
  M.~Czakon,
  Comput.\ Phys.\ Commun.\  {\bf 175}, 559 (2006)
  [arXiv:hep-ph/0511200].

\bibitem{diffeqs}
  T.~Gehrmann and E.~Remiddi,
  Nucl.\ Phys.\ B {\bf 580}, 485 (2000)
  [arXiv:hep-ph/9912329].

\bibitem{sector1}
  T.~Binoth and G.~Heinrich,
  Nucl.\ Phys.\ B {\bf 585}, 741 (2000)
  [arXiv:hep-ph/0004013].

\bibitem{sector2}
  T.~Binoth and G.~Heinrich,
  Nucl.\ Phys.\ B {\bf 680}, 375 (2004)
  [arXiv:hep-ph/0305234].

\bibitem{muon}
  C.~Anastasiou, K.~Melnikov and F.~Petriello,
  arXiv:hep-ph/0505069.


\bibitem{air}
  C.~Anastasiou and A.~Lazopoulos,
  JHEP {\bf 0407}, 046 (2004)
  [arXiv:hep-ph/0404258].

\bibitem{ale-reduction}
A. Daleo, unpublished.



\bibitem{Bonciani:2003te}
  R.~Bonciani, P.~Mastrolia and E.~Remiddi,
  Nucl.\ Phys.\ B {\bf 661}, 289 (2003)
  [Erratum-ibid.\ B {\bf 702}, 359 (2004)]
  [arXiv:hep-ph/0301170].

\bibitem{Bonciani:2003hc}
  R.~Bonciani, P.~Mastrolia and E.~Remiddi,
  Nucl.\ Phys.\ B {\bf 690}, 138 (2004)
  [arXiv:hep-ph/0311145].

\bibitem{Davydychev:2003mv}
  A.~I.~Davydychev and M.~Y.~Kalmykov,
  Nucl.\ Phys.\ B {\bf 699}, 3 (2004)
  [arXiv:hep-th/0303162].

\bibitem{Fleischer:2004vb}
  J.~Fleischer, O.~V.~Tarasov and V.~O.~Tarasov,
  Phys.\ Lett.\ B {\bf 584}, 294 (2004)
  [arXiv:hep-ph/0401090].


\bibitem{Davydychev:2000na}
   A.~I.~Davydychev and M.~Y.~Kalmykov,
   Nucl.\ Phys.\ B {\bf 605} (2001) 266
   [arXiv:hep-th/0012189].

\bibitem{Broadhurst:1993mw}
   D.~J.~Broadhurst, J.~Fleischer and O.~V.~Tarasov,
   Z.\ Phys.\ C {\bf 60}, 287 (1993)
   [arXiv:hep-ph/9304303].

\bibitem{Kalmykov:2000qe}
   M.~Y.~Kalmykov and O.~Veretin,
   Phys.\ Lett.\ B {\bf 483}, 315 (2000)
   [arXiv:hep-th/0004010].


\bibitem{analcont1}
  T.~Gehrmann and E.~Remiddi,
  Comput.\ Phys.\ Commun.\  {\bf 141}, 296 (2001)
  [arXiv:hep-ph/0107173].


\bibitem{Harlander:2005rq}
R.~Harlander and P.~Kant,
JHEP {\bf 0512}, 015 (2005)
[arXiv:hep-ph/0509189].

\bibitem{Spira:1993bb}
M.~Spira, A.~Djouadi, D.~Graudenz and P.~M.~Zerwas,
Phys.\ Lett.\ B {\bf 318}, 347 (1993).



\bibitem{Dawson:1996xz}
  S.~Dawson, A.~Djouadi and M.~Spira,
  Phys.\ Rev.\ Lett.\  {\bf 77}, 16 (1996)
  [arXiv:hep-ph/9603423].

\bibitem{hp2spira}
Muhlleitner, M. Spira, presented during the HP$^2$ workshop, ETH Z\"urich, 
Sept. 2006  

\bibitem{Remiddi:1999ew}
  E.~Remiddi and J.~A.~M.~Vermaseren,
  Int.\ J.\ Mod.\ Phys.\ A {\bf 15} (2000) 725
  [arXiv:hep-ph/9905237].


\bibitem{Tkachov:1981wb}
  F.~V.~Tkachov,
  Phys.\ Lett.\ B {\bf 100}, 65 (1981).

\bibitem{Chetyrkin:1981qh}
  K.~G.~Chetyrkin and F.~V.~Tkachov,
  Nucl.\ Phys.\ B {\bf 192}, 159 (1981).

\bibitem{Kotikov:1990kg}
  A.~V.~Kotikov,
  Phys.\ Lett.\ B {\bf 254}, 158 (1991).

\bibitem{Kotikov:1991hm}
  A.~V.~Kotikov,
  Phys.\ Lett.\ B {\bf 259}, 314 (1991).

\bibitem{Kotikov:1991pm}
  A.~V.~Kotikov,
  Phys.\ Lett.\ B {\bf 267}, 123 (1991).

\bibitem{Remiddi:1997ny}
  E.~Remiddi,
  Nuovo Cim.\ A {\bf 110}, 1435 (1997)
  [arXiv:hep-th/9711188].

\bibitem{Caffo:1998yd}
  M.~Caffo, H.~Czyz, S.~Laporta and E.~Remiddi,
  Acta Phys.\ Polon.\ B {\bf 29}, 2627 (1998)
  [arXiv:hep-th/9807119];
  M.~Caffo, H.~Czyz, S.~Laporta and E.~Remiddi,
  Nuovo Cim.\ A {\bf 111}, 365 (1998)
  [arXiv:hep-th/9805118].

\bibitem{Maitre:2005uu}
  D.~Maitre,
  Comput.\ Phys.\ Commun.\  {\bf 174}, 222 (2006)
  [arXiv:hep-ph/0507152].

\bibitem{Kalmykov:2005hb}
  M.~Y.~Kalmykov,
  Nucl.\ Phys.\ B {\bf 718} (2005) 276
  [arXiv:hep-ph/0503070].

\bibitem{Kalmykov:2004xg}
  M.~Y.~Kalmykov and A.~Sheplyakov,
  Comput.\ Phys.\ Commun.\  {\bf 172} (2005) 45
  [arXiv:hep-ph/0411100].


\bibitem{Nason:1989zy}
  P.~Nason, S.~Dawson and R.~K.~Ellis,
  Nucl.\ Phys.\ B {\bf 327}, 49 (1989)
  [Erratum-ibid.\ B {\bf 335}, 260 (1990)].

\bibitem{Rodrigo:1993hc}
  G.~Rodrigo and A.~Santamaria,
  Phys.\ Lett.\ B {\bf 313}, 441 (1993)
  [arXiv:hep-ph/9305305].


\bibitem{Djouadi:1991tk}
  A.~Djouadi, M.~Spira and P.~M.~Zerwas,
  Phys.\ Lett.\ B {\bf 264}, 440 (1991).



  


\end{thebibliography}
\end{document}